\documentclass[aps,prd,twocolumn,groupedaddress,amsfonts,amssymb,
amsmath,showpacs,floatfix]{revtex4-1}
\usepackage{dcolumn}
\usepackage{multirow}
\usepackage{verbatim}
\usepackage{graphicx}
\usepackage{color}
\usepackage{hyperref}
\newcommand{\fortran}{\texttt{{\normalsize F}{\footnotesize ORTRAN}}~}

\newcommand{\hippy}{\texttt{{\normalsize H}{\footnotesize I}{\normalsize
PP}{\footnotesize Y}}~}
\newcommand{\hpsrc}{\texttt{{\normalsize HP}{\footnotesize SRC}}~}
\newcommand{\mathematica}{\texttt{Mathematica}~}

\newcommand{\als}{\alpha_s}

\newcommand{\msb}{{\overline{\rm MS}}}

\begin{document}

%\preprint{}

\title{Matching lattice and continuum four-fermion operators with
nonrelativistic QCD and highly improved staggered quarks}


\author{Christopher \surname{Monahan}}
%\thanks{}
\affiliation{Physics Department, College of William and Mary,
Williamsburg, Virginia 23187, USA}
\author{Elvira \surname{G\'amiz}}
\affiliation{CAFPE and Departamento de F\'isica Te\'orica y del Cosmos, 
Universidad de Granada, E-18071 Granada, Spain}
\author{Ron \surname{Horgan}}
\affiliation{DAMTP, Centre for Mathematical Sciences, University of
Cambridge, Cambridge, CB3 0WA, UK}
\author{Junko \surname{Shigemitsu}}
\affiliation{Physics Department, The Ohio State University, Columbus,
Ohio 43210, USA}

%used with the \author comman(d).
\collaboration{HPQCD collaboration}
\noaffiliation

%\date{\today}

\begin{abstract}
We match continuum and lattice heavy-light four-fermion operators at one 
loop in perturbation theory. For the 
heavy quarks we use nonrelativistic QCD and for the 
massless light quarks the highly 
improved staggered quark action. We include the full set of $\Delta B=2$ 
operators 
relevant to neutral $B$ mixing both within and beyond the standard model and 
match through order $\alpha_s$, $\Lambda_{\mathrm{QCD}}/M_b$, and 
$\alpha_s/(aM_b)$.
\end{abstract}

%\keywords{}

\maketitle

\section{Introduction}

Despite intense experimental and theoretical effort, there have been no 
observations of beyond the standard model (BSM) particles. Direct detection at 
high energy collider experiments is 
not, however, 
the only way to uncover evidence for new physics. Indirect detection through 
high-precision measurements at relatively 
low energies is also possible. At low energies, new physics appears through 
quantum loop 
effects, which can probe energy scales far greater than those available at 
current high energy experiments, such as at the Large Hadron Collider. 
Detecting such loop effects requires precise theoretical 
predictions of standard model physics with which to compare
experimental data. A related approach is to study the unitarity of the 
Cabibbo-Kobayashi-Maskawa (CKM) quark mixing matrix. In the 
standard model, the CKM 
matrix is unitary and deviations from unitarity could indicate the presence of 
new physics. Multiple, independent determinations overconstrain the CKM 
parameters, usually expressed in terms of ``unitarity triangles''. 

Heavy quark flavor physics is one area that could be 
particularly sensitive to the effects of heavy BSM particles. In particular, 
neutral $B$ meson mixing, which is both loop suppressed and 
CKM suppressed, provides a promising 
avenue for new physics searches. In the past decade there have been extensive 
experimental studies of neutral $B$ meson mixing and $B$ decays from the CDF 
\cite{Aaltonen:2012ie,CDF:2011af}, D0 
\cite{Abazov:2013uma,Abazov:2012hha,Abazov:2010hj}, and most recently, LHCb 
\cite{Aaij:2013gja,Aaij:2012nt} collaborations. Some of these results have 
exposed a 2-3$\sigma$ discrepancy between certain standard model predictions 
and measurements \cite{Lunghi:2010gv,Abazov:2013uma,Abazov:2010hj}. In 
addition, recent CKM unitarity triangle fits 
hint at the presence of BSM physics, with some fits favoring new 
physics contributions in the neutral $B$ mixing sector 
\cite{Lenz:2012az,Laiho:2011nz,Lenz:2010gu,Lunghi:2009ke}.

Neutral $B$ meson mixing is characterized by the mass and decay width
differences between the ``heavy'' and ``light'' mass eigenstates, 
which are admixtures of quark flavor eigenstates. The mass difference, $\Delta 
M_q=M_H-M_L$, is equivalent to the oscillation frequency of a neutral $B_q$ 
meson with light quark species $q$. Theoretical studies of neutral $B$ meson 
mixing 
employ effective Hamiltonians that incorporate four-fermion operators. Matrix 
elements of these operators characterize the nonperturbative quantum 
chromodynamics (QCD) behavior of the mixing process and these matrix elements 
must be determined with a precision sufficient to confront experimental data 
with stringent tests. Precise \emph{ab initio} calculations of nonperturbative 
QCD effects require lattice QCD.

The scope of neutral $B$ meson mixing calculations on the lattice has been quite
extensive and several lattice collaborations have produced results with up/down 
and strange quarks in the sea 
\cite{Aoki:2014nga,Chang:2013gla,Ishikawa:2013faa,Bazavov:2012zs,Gamiz:2009ku}. 
The HPQCD 
collaboration is currently carrying out nonperturbative calculations that 
incorporate the effects of up/down, strange, and charm quarks in the sea for 
the first time \cite{inprep}.

The gauge ensembles that are currently available have a 
lattice spacing too large to accommodate heavy quarks 
directly at the physical $b$ quark mass. Lattice calculations are 
therefore generally carried out using an effective theory for the heavy quark 
fields, such as heavy quark effective theory (HQET) or nonrelativistic QCD 
(NRQCD). Effective theories on the lattice must be related to continuum QCD to 
extract physically meaningful results. In this paper we determine the one loop 
matching coefficients required to relate lattice matrix elements of $\Delta 
B=2$ operators, constructed using the highly improved staggered quark (HISQ) 
and NRQCD actions, to the corresponding matrix elements in continuum QCD. We 
match through ${\cal 
O}(\alpha_s,\Lambda_{QCD}/M_b,\alpha_s/(aM_b))$ and include 
``subtracted'' dimension-seven operators, which remove power law divergences at 
${\cal O}(\alpha_s/(aM))$, only at tree level.

Our calculation extends the work of \cite{Gamiz:2008sk} to include massless 
HISQ light quarks and is a significant step in the HPQCD collaboration's 
program 
to determine improvement and matching coefficients for lattice NRQCD at one 
loop \cite{Dowdall:2011wh,Hammant:2013sca,Monahan:2012dq}. These matching 
calculations are an integral component of the HPQCD collaboration's precision 
$B$ physics effort. Here we largely follow the notation of \cite{Gamiz:2008sk} 
for 
consistency and to enable easy comparison with that paper. A similar matching 
calculation for a restricted range of $\Delta B=2$ operators in NRQCD was 
carried out in \cite{Hashimoto:2000eh}. Matching calculations for static heavy 
quarks with a range of light quark actions were undertaken in 
\cite{Ishikawa:1998rv} and more recently in 
\cite{Loktik:2006kz,Christ:2007cn,Ishikawa:2011dd}. A preliminary discussion of 
${\cal O}(1/M_b)$ operators in HQET was presented in \cite{Papinutto:2013cra}. 
We provide full details of the extraction of the lattice NRQCD mixing 
coefficients, which does not appear in the literature.

In the next section we discuss four-fermion operators in continuum QCD 
and on the lattice. We then describe the matching procedure that relates 
the matrix elements of these operators. In Sec.~\ref{sec:mixing} we 
detail the calculation of the lattice mixing coefficients. We present our 
results for the
mixing parameters from heavy-light four-fermion operators through order 
$\alpha_s$, $\Lambda_{\mathrm{QCD}}/M_b$, and $\alpha_s/(aM_b)$ in Sec.~
\ref{sec:matchresults}. We conclude with a summary in
Sec.~\ref{sec:summary}. In the Appendix we provide some 
details of the continuum calculations entering the matching procedure. We 
discuss two different NDR-$\msb$ schemes that have been used in the literature 
for the renormalization of the standard model $\Delta B=2$ operators $Q2$ 
and $Q3$, and we correct two errors in Eqs.~(B9) and (B10) of Ref.~
\cite{Gamiz:2008sk}.

\section{Four-fermion operators}
\subsection{In continuum QCD}

There are three dimension-six, $\Delta B=2$ operators that are relevant 
to neutral $B$ meson mixing in the standard model:
\begin{align}
Q1 = {} & \left(\overline{\Psi}^i_b\gamma^\mu P_L \Psi^i_q\right)
\left(\overline{\Psi}^j_b\gamma_\mu P_L \Psi^j_q\right) , \label{eq:q1} \\
Q2 = {} & \left(\overline{\Psi}^i_b P_L \Psi^i_q\right)
\left(\overline{\Psi}^j_b P_L \Psi^j_q\right) , \label{eq:q2} \\
Q3 = {} & \left(\overline{\Psi}^i_b P_L \Psi^j_q\right)
\left(\overline{\Psi}^j_b P_L \Psi^i_q\right). \label{eq:q3} 
\end{align}
Here the subscript on the QCD fields, $\Psi$ and $\overline{\Psi}$, denotes the 
quark species: $b$ for bottom quarks and $q$ for down or strange quarks, which 
we take to be massless. The superscripts $i$ and $j$ are color indices and 
$P_{R,L}=(1\pm\gamma_5)$ are right- and left-handed projectors. Operator $Q1$ 
determines the mass difference $\Delta M_q$ in the standard model and all three 
are useful in studies of the width difference $\Delta \Gamma_q$.

BSM physics can be parametrized by a $\Delta 
B=2$ effective Hamiltonian, which incorporates two further 
independent operators,
\begin{align}
Q4 = {} & \left(\overline{\Psi}^i_b P_L \Psi^i_q\right)
\left(\overline{\Psi}^j_b P_R \Psi^j_q\right) , \label{eq:q4} \\
Q5 = {} & \left(\overline{\Psi}^i_b P_L \Psi^j_q\right)
\left(\overline{\Psi}^j_b P_R \Psi^i_q\right). \label{eq:q5} 
\end{align}
Collectively these five operators are known as the ``SUSY basis of 
operators'' in the literature \cite{Gabbiani:1996hi}. We simplify intermediate 
stages of 
the matching calculation by introducing two extra operators,
\begin{align}
Q6 = {} & \left(\overline{\Psi}^i_b \gamma_\mu P_L \Psi^i_q\right)
\left(\overline{\Psi}^j_b \gamma^\mu P_R \Psi^j_q\right) , \label{eq:q6} \\
Q7 = {} & \left(\overline{\Psi}^i_b \gamma_\mu P_L \Psi^j_q\right)
\left(\overline{\Psi}^j_b \gamma^\mu P_R \Psi^i_q\right). \label{eq:q7} 
\end{align}
Matrix elements of these operators are related to matrix elements of $Q5$ 
and $Q4$ via Fierz relations, so that, as one would expect, $Q6$ and $Q7$ are 
not independent operators.

Matching calculations in perturbation theory are generally carried out by 
considering scattering between external quark (or gluon) states. For the case 
of $\Delta B = 2$ operators, we consider scattering from an incoming state 
consisting of a heavy antiquark and a light quark to an outgoing state of a 
heavy quark and light antiquark. We write these states symbolically by
\begin{equation}
\left|\mathrm{in}\right\rangle = 
\big|\overline{Q}^B;q^C\big\rangle,\quad\mathrm{and}\quad  \left\langle 
\mathrm{out}\right| = \left\langle \overline{q}^A;Q^D\right|,
\end{equation}
where the superscripts are color indices. The corresponding external Dirac 
spinors are $u_q$ and $v_q$ for the incoming light quark and outgoing light 
antiquark and $\overline{u}_Q$ and $\overline{v}_Q$ for the outgoing heavy 
quark and incoming heavy antiquark respectively.

\begin{figure}
\includegraphics[width=0.4\textwidth,keepaspectratio=true]{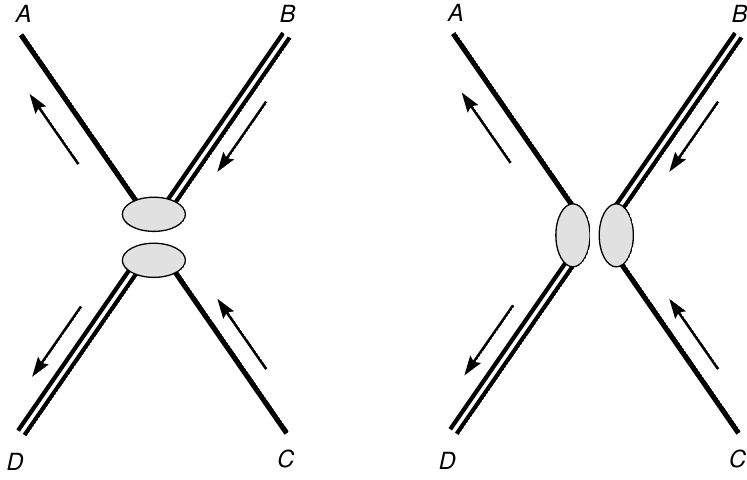}
\caption{\label{fig:tree} 
Tree-level diagrams representing the matrix elements of operators $Q1$, $Q2$, 
$Q4$, and $Q6$. The incoming state is a heavy antiquark and a light 
quark and 
the outgoing state is a heavy quark and a light antiquark. The letters $A$, 
$B$, $C$, and $D$ are color indices and correspond to the conventions of 
Eq.~\protect\eqref{eq:q124tree}.}
\end{figure}

We denote the matrix elements of the operators $Qi$ by
\begin{equation}
\langle Qi \rangle = \left\langle\mathrm{out} \right| Qi 
\left|\mathrm{in}\right\rangle,
\end{equation}
and at tree level $Q1$, $Q2$, $Q4$, and 
$Q6$ are
\begin{align}\label{eq:q124tree}
{} & \left\langle 
 \overline{q}^A;Q^D\right| \left(\overline{\Psi}_b^i\Gamma_1 
\Psi_q^i\right) \left(\overline{\Psi}_b^j\Gamma_2 
\Psi_q^j\right)\big|\overline{Q}^B;q^C\big\rangle_{\mathrm{tree}} \nonumber\\
{} &  = 
\delta_{AB}\delta_{CD}\left[(\overline{u}_Q\Gamma_1u_q)(\overline{v} 
_Q\Gamma_2v_q)+(\overline{u}_Q\Gamma_2u_q)(\overline{v}
_Q\Gamma_1v_q)\right]\nonumber\\
{} & \quad - 
\delta_{AD}\delta_{BC}
\left[(\overline{u}_Q\Gamma_1v_q)(\overline{v} 
_Q\Gamma_2u_q)+(\overline{u}_Q\Gamma_2v_q)(\overline{v}
_Q\Gamma_1u_q)\right],
\end{align}
which we represent diagrammatically in Fig.~\ref{fig:tree}. The Dirac 
operators $\Gamma_{1,2}$ represent the operator insertions corresponding to 
Eqs.~\eqref{eq:q1} to \eqref{eq:q7}.
For matrix elements of $Q3$, $Q5$, and $Q7$, we have instead
\begin{align}\label{eq:q357tree}
{} & \left\langle 
 \overline{q}^A;Q^D\right| \left(\overline{\Psi}_b^i\Gamma_1 
\Psi_q^j\right) \left(\overline{\Psi}_b^j\Gamma_2 
\Psi_q^i\right)\big|\overline{Q}^B;q^C\big\rangle_{\mathrm{tree}} \nonumber\\
{} &  = 
\delta_{AD}\delta_{CB}\left[(\overline{u}_Q\Gamma_1u_q)(\overline{v} 
_Q\Gamma_2v_q)+(\overline{u}_Q\Gamma_2u_q)(\overline{v}
_Q\Gamma_1v_q)\right]\nonumber\\
{} & \quad - 
\delta_{AB}\delta_{CD}
\left[(\overline{u}_Q\Gamma_1v_q)(\overline{v} 
_Q\Gamma_2u_q)+(\overline{u}_Q\Gamma_2v_q)(\overline{v}
_Q\Gamma_1u_q)\right].
\end{align}

Radiative corrections induce mixing between the four-fermion operators, 
which we write as
\begin{equation}\label{eq:qcdcij}
\langle Qi \rangle^{\overline{MS}} = \langle Qi \rangle_{\mathrm{tree}} 
+\alpha_s c_{ij}\langle Qj \rangle_{\mathrm{tree}}^{(0)},
\end{equation}
where the superscript $(0)$ denotes matrix elements constructed using spinors 
that obey
\begin{equation}
\overline{u}_Q\gamma_0 = \overline{u}_Q,\quad \mathrm{and}\quad 
\overline{v}_Q\gamma_0 = - \overline{v}_Q,
\end{equation}
in order to match to the effective theory. In 
principle the product $c_{ij}\langle Qj \rangle_{\mathrm{tree}}^{(0)}$ is a sum 
over all operators $Qj$ that mix with $Qi$. In practice, however, only two such 
operators appear: for example, for $Q1$ we have
\begin{equation}
\langle Q1 \rangle^{\overline{MS}} = \langle Q1 \rangle_{\mathrm{tree}}
+\alpha_s c_{11}\langle Q1 \rangle_{\mathrm{tree}}^{(0)}
+\alpha_s c_{12}\langle Q2 \rangle_{\mathrm{tree}}^{(0)}.
\end{equation}
In the following, we leave this sum implicit.

\subsection{On the lattice}

In the effective theory formalism of NRQCD, the heavy quarks and 
antiquarks are treated as distinct quark species. We separate the quark fields 
that create 
heavy quarks, which we denote $\overline{\Psi}_Q$, from the fields that 
annihilate heavy antiquarks, which we represent by 
$\overline{\Psi}_{\overline{Q}}$.

The two-component heavy quark field is obtained from the four-component QCD 
quark field, $\overline{\Psi}_b$, via the Foldy-Wouthuysen-Tani transformation 
(see, for example, \cite{Itzykson:1980rh}),
\begin{equation}
\overline{\Psi}_b = 
\overline{\Psi}_Q\left(1+\frac{1}{2M}\boldsymbol{\gamma}\cdot\overleftarrow{
\nabla
}+{\cal O}(1/M^2)\right),
\end{equation}
where the arrow indicates that the derivative acts on the heavy quark field to 
the left. We insert this expansion into the four-fermion operators of Eqs.~
\eqref{eq:q1} to \eqref{eq:q7} to determine the appropriate NRQCD operators. We 
see immediately that, at leading order in $1/M$, we need 
operators of the form
\begin{equation}
\widehat{Q}i = 
\left(\overline{\Psi}_Q\Gamma_1\Psi_q\right)\left(\overline{\Psi}
_{\overline{Q}}\Gamma_2\Psi_q\right)+\left(\overline{\Psi}
_{\overline{Q}}\Gamma_1\Psi_q\right)\left(\overline{\Psi}
_Q\Gamma_2\Psi_q\right).
\end{equation}

We obtain the ${\cal O}(\Lambda_{\mathrm{QCD}}/M)$ corrections by introducing 
the operators
\begin{align}
\widehat{Q}i1 = {} & \frac{1}{2M}\Big[
\left(\overrightarrow{\nabla}\overline{\Psi}
_Q\cdot \boldsymbol{\gamma}\Gamma_1\Psi_q\right)\left(\overline { \Psi }
_{\overline{Q}}\Gamma_2\Psi_q\right) \nonumber\\
{} & \quad +\left(\overline{\Psi}
_Q\Gamma_1\Psi_q\right)\left(\overrightarrow{\nabla}
\overline{\Psi}_{\overline{Q}}\cdot \boldsymbol{\gamma}
\Gamma_2\Psi_q\right) \nonumber\\
{} & \quad + \left(\overrightarrow{\nabla}\overline{\Psi}
_{\overline{Q}}\cdot 
\boldsymbol{\gamma}\Gamma_1\Psi_q\right)\left(\overline{\Psi}
_Q\Gamma_2\Psi_q\right)\nonumber\\
{} & \quad + \left(\overline{\Psi}
_{\overline{Q}}\Gamma_1\Psi_q\right)\left(\overrightarrow{\nabla}\overline{
\Psi}
_Q\cdot 
\boldsymbol{\gamma}\Gamma_2\Psi_q\right)
\Big].
\end{align}

We denote the matrix elements of the effective theory by
\begin{equation}
\langle \widehat{Q}i \rangle = \left\langle\mathrm{out} \right| \widehat{Q}i 
\left|\mathrm{in}\right\rangle,\quad \mathrm{and}\quad \langle \widehat{Q}i1 
\rangle = \left\langle\mathrm{out} \right| \widehat{Q}i1
\left|\mathrm{in}\right\rangle,
\end{equation}
where now the ``in'' and ``out'' states are understood to be an incoming NRQCD 
antiquark and HISQ quark and an outgoing NRQCD quark and HISQ antiquark, 
respectively. Radiative corrections induce mixing between these operators, with 
mixing coefficients $c_{ij}^{\mathrm{latt}}$, and we obtain
\begin{equation}\label{eq:lattcij}
\langle \widehat{Q}i \rangle
= \langle \widehat{Q}i 
\rangle_{\mathrm{tree}}^{(0)}+\alpha_s c_{ij}^{\mathrm{latt}}\langle 
\widehat{Q}j 
\rangle_{\mathrm{tree}}^{(0)},
\end{equation}
and similarly
\begin{equation}\label{eq:lattcij1}
\langle \widehat{Q}i1 \rangle
= \langle \widehat{Q}i1 
\rangle_{\mathrm{tree}}^{(0)}+\alpha_s\zeta_{ij}^{\mathrm{latt}}\langle 
\widehat{Q}j 
\rangle_{\mathrm{tree}}^{(0)}.
\end{equation}
We ignore the one loop corrections to $\langle \widehat{Q}i1 
\rangle_{\mathrm{tree}}^{(0)}$, which only arise at ${\cal 
O}(\alpha_s\Lambda_{\mathrm{QCD}}/M_b)$ in the matching procedure. 

As discussed in more detail in \cite{Gamiz:2008sk}, the mixing coefficients 
$\zeta_{ij}^{\mathrm{latt}}$ describe the ``mixing down'' of dimension-seven 
operators $\widehat{Q}i1$ onto dimension-six operators $\widehat{Q}j$.

In the next section we outline the matching procedure before describing the 
calculation of the lattice mixing coefficients.

\section{\label{sec:matching}The Matching Procedure}

We now relate the matrix elements of the NRQCD-HISQ operators, which ultimately 
will be determined nonperturbatively on the lattice, to the matrix elements of 
QCD operators in the $\overline{MS}$ scheme. In other words, we wish to relate 
Eqs.~\eqref{eq:lattcij} and \eqref{eq:lattcij1} to Eq.~
\eqref{eq:qcdcij}.

We first expand the QCD matrix element $\langle Qi \rangle_{\mathrm{tree}}$ in 
Eq.~\eqref{eq:qcdcij} in powers of the inverse heavy quark mass:
\begin{equation}
\langle Qi \rangle_{\mathrm{tree}} = \langle Qi 
\rangle_{\mathrm{tree}}^{(0)}+\langle Qi1 \rangle_{\mathrm{tree}}^{(0)}.
\end{equation}
Thus the QCD matrix element becomes
\begin{equation}\label{eq:qcdtmp}
\langle Qi \rangle^{\overline{MS}} = \langle Qi 
\rangle_{\mathrm{tree}}^{(0)}+\langle Qi1 
\rangle_{\mathrm{tree}}^{(0)}
+\alpha_s c_{ij}\langle Qj \rangle_{\mathrm{tree}}^{(0)} .
\end{equation}

Our aim is to write the QCD matrix element in terms of the NRQCD-HISQ matrix 
elements. Therefore we need to reexpress the tree-level matrix elements 
$\langle Qi \rangle_{\mathrm{tree}}^{(0)}$ and $\langle Qi1 
\rangle_{\mathrm{tree}}^{(0)}$ in terms of the matrix elements on the lattice. 
To achieve this, we invert Eqs.~\eqref{eq:lattcij} and \eqref{eq:lattcij1} 
to obtain
\begin{equation}
\langle \widehat{Q}i 
\rangle_{\mathrm{tree}}^{(0)} = \langle \widehat{Q}i \rangle 
-\als c_{ij}^{\mathrm{latt}}\langle \widehat{Q}j \rangle,
\end{equation}
and
\begin{equation}
\langle \widehat{Q}i1 
\rangle_{\mathrm{tree}}^{(0)} = \langle \widehat{Q}i1 \rangle - 
\alpha_s\zeta_{ij}^{\mathrm{latt}}\langle 
\widehat{Q}j 
\rangle.
\end{equation}

Using
\begin{equation}
\langle \widehat{Q}i 
\rangle_{\mathrm{tree}}^{(0)} = \langle Qi 
\rangle_{\mathrm{tree}}^{(0)},\quad \mathrm{and}\quad \langle \widehat{Q}i1 
\rangle_{\mathrm{tree}}^{(0)} = \langle Qi1
\rangle_{\mathrm{tree}}^{(0)},
\end{equation}
we can now plug these results into Eq.~\eqref{eq:qcdtmp} to find
\begin{align}
\langle Qi \rangle^{\overline{MS}} = {} & 
\left[1+\alpha_s\rho_{ii}\right]\langle \widehat{Q}i \rangle  + 
\alpha_s\rho_{ij}\langle 
\widehat{Q}j \rangle+ \langle \widehat{Q}i1 \rangle  \nonumber\\
{}& \quad - 
\alpha_s\zeta_{ij}^{\mathrm{latt}}\langle 
\widehat{Q}j 
\rangle  + {\cal O}(\alpha_s^2,\alpha_s\Lambda_{\mathrm{QCD}}/M),
\end{align}
where the matching coefficients, $\rho_{ij}$, are given by
\begin{equation}
\rho_{ij} = c_{ij}-c^{\mathrm{latt}}_{ij}.
\end{equation}

We now define the ``subtracted'' matrix elements, which remove power law 
divergences at ${\cal O}(\alpha_s/(aM))$ \cite{Gamiz:2008sk}, as 
\begin{equation}
\langle \widehat{Q}i1 \rangle_{\mathrm{sub}} = \langle \widehat{Q}i1 \rangle - 
\alpha_s\zeta_{ij}^{\mathrm{latt}}\langle 
\widehat{Q}j 
\rangle,
\end{equation}
so that our final expression is
\begin{align}
\langle Qi \rangle^{\overline{MS}} = {} & 
\langle \widehat{Q}i \rangle  + 
\alpha_s\rho_{ij}\langle 
\widehat{Q}j \rangle+ \langle \widehat{Q}i1 \rangle_{\mathrm{sub}} \nonumber\\
{}& \qquad  + {\cal 
O}(\alpha_s^2,\alpha_s\Lambda_{\mathrm{QCD}}/M).
\end{align}
For a more comprehensive discussion of power 
law divergences in lattice NRQCD see \cite{Collins:2000ix} and 
\cite{Morningstar:1997ep}.

\section{\label{sec:mixing}Evaluation of lattice mixing coefficients}

Complete details of the lattice actions 
used in our matching 
procedure were given in \cite{Monahan:2012dq} and here we simply summarize the 
relevant information. For the gauge fields we use the Symanzik improved gauge 
action with tree level 
coefficients \cite{Weisz:1982zw,Weisz:1983bn,Curci:1983an,Luscher:1985zq}, 
because radiative improvements 
to the gluon action do not contribute to the matching calculation at one loop 
\cite{Monahan:2012dq}. We include a 
gauge-fixing term and, where possible, we confirm that gauge invariant
quantities are gauge parameter independent by working in
both Feynman and Landau gauges.

We discretize the light quarks using the HISQ action \cite{Follana:2006rc} and 
set the bare light 
quark 
mass to zero. For the heavy quark fields, we 
use the tree-level NRQCD action of \cite{Dowdall:2011wh,Monahan:2012dq}. We do 
not consider the effects of radiative improvement of the NRQCD action, which 
are not required for our one loop calculation.

Our results were obtained using two independent methods: 
with the automated lattice perturbation theory
routines \hippy and \hpsrc \cite{Hart:2004bd,Hart:2009nr}; and with 
\mathematica 
and 
\fortran 
routines developed for earlier matching calculations 
\cite{Dalgic:2003uf,Monahan:2012dq}. We described both of these methods in 
detail in \cite{Monahan:2012dq}.

We undertook a number of checks of our results. We reproduced the results of 
\cite{Gamiz:2008sk} with NRQCD heavy quarks and AsqTad light quarks to test 
the automated lattice perturbation theory routines. In many 
cases, we established that gauge invariant
quantities, such as the mass renormalization, are gauge parameter
independent by working in both Feynman and Landau gauges. Furthermore, we
carried out several diagram specific checks, which we discuss in more detail in 
the next subsections.

Finally, we confirmed that infrared
divergent parameters, such as the wavefunction renormalization and certain 
matching parameters, exhibited
the correct continuum-like behavior. 

As with the heavy-light current matching results of 
\cite{Monahan:2012dq}, we believe that these two methods are 
sufficiently independent that agreement between these methods provides a 
stringent check of our results.

\subsection{Dimension-six operators}

\begin{figure}
\includegraphics[width=0.48\textwidth,keepaspectratio=true]{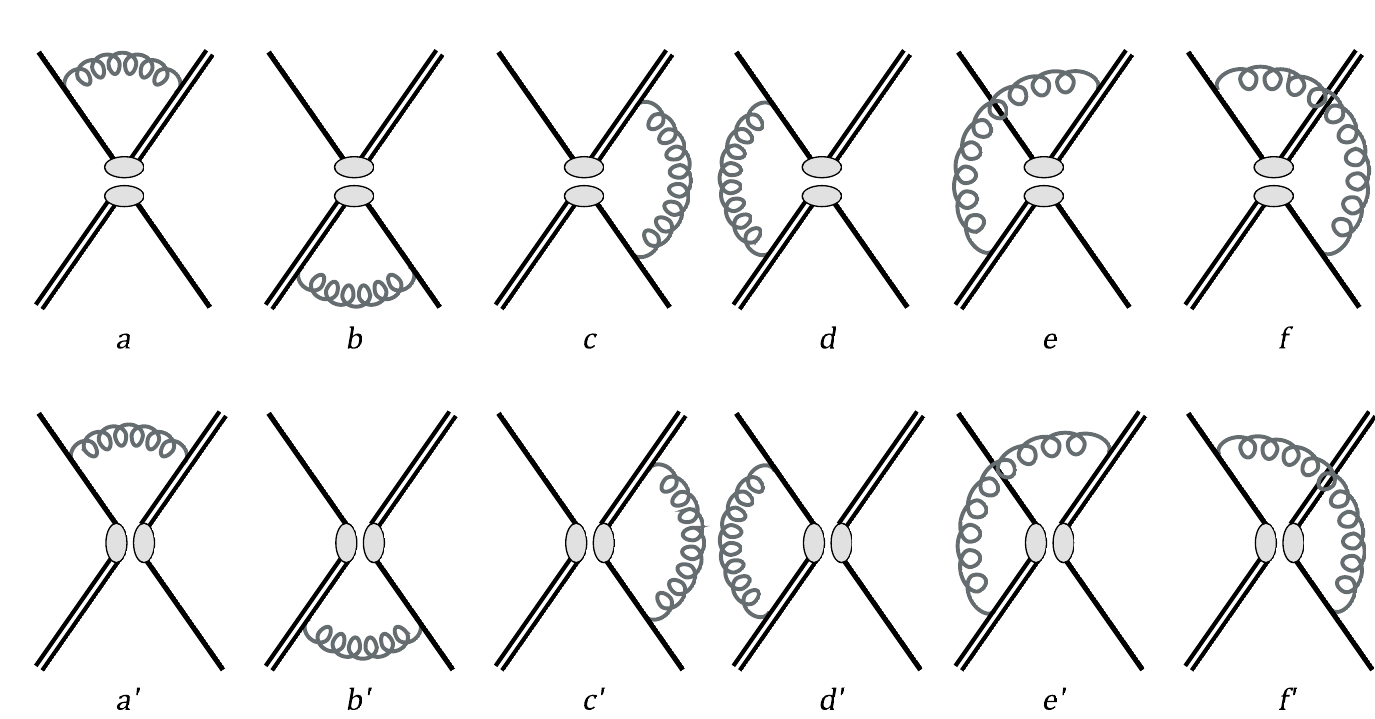}
\caption{\label{fig:oneloop} 
One loop diagrams representing the corrections to matrix elements of 
the operators $Qi$. The external states are those of 
Fig.~\protect\ref{fig:tree} and 
Eq.~\protect\eqref{eq:q124tree}.}
\end{figure}

The spinor structures corresponding to the one loop 
contributions to the matrix elements of the dimension-six operators of Eq.~
\eqref{eq:lattcij} can be written schematically as the product of two spinor 
bilinears, each with some particular Lorentz and color structure specified by 
the precise contribution in question. We illustrate the corresponding Feynman 
diagrams in Fig.~\ref{fig:oneloop}.
This idea schematically generalizes the tree-level results of Eqs.~
\eqref{eq:q124tree} and \eqref{eq:q357tree}.

With this schematic in mind, we 
can break down the diagrams of Fig.~\ref{fig:oneloop} into two types: 
those diagrams in which a gluon propagator connects each spinor bilinear and 
those without such a propagator connection. It is straightforward to 
recognize that diagrams (a), (b), (c$^\prime$), and (d$^\prime$) of Fig.~
\ref{fig:oneloop} fall into the latter 
category and all others into the former. In the following, we focus the 
discussion on the determination of mixing coefficients for $\widehat{Q}1$, 
$\widehat{Q}2$, $\widehat{Q}4$, 
and $\widehat{Q}6$. We discuss $\widehat{Q}3$, $\widehat{Q}5$, and 
$\widehat{Q}7$ at the end of this 
subsection.

\subsubsection{Diagrams (a) and (b)}

Diagrams (a) and (b) are the most straightforward to compute, 
since we can separate the spinor bilinears. Diagrams (c$^\prime$) and 
(d$^\prime$) are similarly straightforward, but only contribute to 
$\widehat{Q}3$, $\widehat{Q}5$, and $\widehat{Q}7$, which we discuss later. The 
contribution to $\widehat{Q}1$ from diagram (a) is given by
\begin{equation}
 \mathrm{(a)} = \frac{4}{3}\delta_{AB}\delta_{CD} \Big(\overline{u}_Q\gamma^\mu 
P_L u_q\Big)\Big(\overline{v}_{\overline{Q}}{\cal V}_\mu  v_q\Big),
\end{equation}
where ${\cal V}_\mu$ represents the one loop vertex correction to the 
heavy-light quark bilinear $\overline{v}_{\overline{Q}}\gamma_\mu P_L  v_q$:
\begin{equation}\label{eq:vmu}
{\cal V}_\mu = 
V_{\overline{Q}\overline{Q}g}^\nu 
\,G_{\overline{Q}}\,\gamma_\mu \,P_L \,G_q \, V_{qqg}^\sigma\, K_{\nu\sigma}.
\end{equation}
Here the $V^\nu$ represent the quark-quark-gluon vertices, $G_{\overline{Q}}$ 
the heavy antiquark propagator and $G_q$ the quark propagator, and $ 
K_{\nu\sigma}$ the gluon propagator. Note that, for the other operators in the 
SUSY basis, there is no occurrence of $\gamma_\mu$ in the operator insertions 
and consequently diagram (a) takes the form
\begin{equation}
 \mathrm{(a)} = \frac{4}{3}\delta_{AB}\delta_{CD} \Big(\overline{u}_Q 
P_{L,R} u_q\Big)\Big(\overline{v}_{\overline{Q}}{\cal V} \, v_q\Big),
\end{equation}
where
\begin{equation}
{\cal V} = 
V_{\overline{Q}\overline{Q}g}^\nu 
\,G_{\overline{Q}}\, \,P_{L,R} \,G_q \, V_{qqg}^\sigma\, K_{\nu\sigma}.
\end{equation}

We have chosen a specific combination of external 
colors that isolates the contribution proportional 
to the spinor bilinears $\overline{u}_Q\gamma^\mu P_L 
u_q$ and $\overline{v}_{\overline{Q}}\gamma^\mu P_L   v_q$ [compare to Eq.~
\eqref{eq:q124tree}], with color factor $(4/3)\delta_{AB}\delta_{CD}$. We could 
equally have chosen to isolate the spinor structure proportional 
to $\overline{u}_Q\gamma^\mu P_L 
v_q$ and $\overline{v}_{\overline{Q}}\gamma^\mu P_L   u_q$ with color factor 
$(-4/3)\delta_{AD}\delta_{BC}$. This choice would have given identical results. 
In 
the following discussion we leave the color factor 
implicit for clarity and always work with the contribution to ${\cal O}1$ 
(analogous 
relations hold for 
the other operators).

We separate out the temporal and spatial components so that, for diagram (a), 
for example, we write
\begin{align}\label{eq:ac0}
 \mathrm{(a)} = {} &  c_0\Big(\overline{u}_Q\gamma^0 P_L 
u_q\Big)\Big(\overline{v}_{\overline{Q}}\gamma^0 P_L 
v_q\Big) \nonumber\\
{} & \quad +\sum_{k=1}^3c_k\Big(\overline{u}_Q\gamma^k P_L 
u_q\Big)\Big(\overline{v}_{\overline{Q}}\gamma^k P_L   v_q\Big).
\end{align}
By symmetry of the spatial directions, the three coefficients $c_k$, for 
$k\in\{1,2,3\}$, are all equal. In terms of the operator mixing of Eq.~
\eqref{eq:lattcij}, we also have
\begin{align}
 \mathrm{(a)} = {} &  c_{11}^{\mathrm{latt}}\Big(\overline{u}_Q\gamma^\mu P_L 
u_q\Big)\Big(\overline{v}_{\overline{Q}}\gamma_\mu P_L 
v_q\Big) \nonumber\\
{} & \quad +c_{12}^{\mathrm{latt}}\Big(\overline{u}_Q P_L 
u_q\Big)\Big(\overline{v}_{\overline{Q}}P_L   v_q\Big).
\end{align}
Therefore, by projecting out the coefficient of each spinor structure in 
Eq.~\eqref{eq:ac0}, we can obtain the mixing coefficients from
\begin{equation}
c_{11}^{\mathrm{latt}} = c_k,\quad \mathrm{and}\quad c_{12}^{\mathrm{latt}} = 
c_k - c_0.
\end{equation}

In the automated lattice perturbation theory routines used in this calculation, 
the result of a generic Feynman diagram integral is expressed as a 
``\verb+spinor+'', which is a derived type specified by the \hpsrc module 
\verb+mod_spinors.F90+ \cite{Hart:2004bd,Hart:2009nr}. The \verb+spinor+ type 
incorporates a 16-element array that specifies the coefficient of each 
element of the Dirac algebra. Therefore, to extract the coefficient of some 
particular Dirac structure, all one needs to do is return the corresponding 
element of the \verb+spinor+ array (external spinors are dropped for 
the purposes of the calculation). 

For example, to determine $c_k$ for diagram (a) 
we extract the coefficient of, say, $\gamma_3$ from the integrated expression 
for the Feynman diagram. This corresponds exactly to the standard continuum 
procedure of multiplying by an appropriate projector and taking the trace, 
which is the method applied in our second, \mathematica/\fortran, approach to 
this calculation.

We applied two sets of cross-checks to our results for these diagrams. First, 
we checked that diagrams (a) and (b) give identical results. Second, we 
confirmed 
that the mixing coefficients were equal to the corresponding heavy-light 
current results of \cite{Monahan:2012dq}:
\begin{equation}
c_{11}^{\mathrm{latt},\,(a)}=\zeta_{11}^{(V_k)},\;
c_{22}^{\mathrm{latt},\,(a)}=\zeta_{11}^{(V_0)},\; 
c_{12}^{\mathrm{latt},\,(a)}=\zeta_{11}^{(V_k)}-\zeta_{11}^{(V_0)}.
\end{equation}
Note that these $\zeta_{11}^{(V_\mu)}$ are not the mixing coefficients of the 
$1/M$ operators described above (which we denote $\zeta_{ij}^{\mathrm{latt}}$), 
but the mixing coefficients of the heavy-light currents described in 
\cite{Monahan:2012dq}.

\subsubsection{Diagrams (c) to (f$^{\,\,\prime}\!$)}

The calculation of the contributions from diagrams (c) to (f$^\prime$) of 
Fig.~\ref{fig:oneloop} proceed along conceptually similar lines, although the 
integrand structure is more complicated.

We will examine two examples of the 
possible spinor structure to illustrate our method. The other diagrams follow 
the same pattern, \emph{mutatis mutandis}.

The contribution to $\widehat{Q}1$ from diagram (c) is given by
\begin{equation}
 \mathrm{(c)} = -\frac{1}{6}\delta_{AB}\delta_{CD} \Big(\overline{u}_Q{\cal 
V}^{(1)\,\mu\nu} u_q\Big)\Big(\overline{v}_{\overline{Q}}{\cal 
V}^{(2)}_{\mu\nu}  v_q\Big),
\end{equation}
where, using the notation described below Eq.~\eqref{eq:vmu},
\begin{equation}\label{eq:vmunu}
{\cal V}^{(1)\,\mu\nu} = 
\gamma^\mu P_L G_q V_{qqg}^\nu, \quad
{\cal V}^{(2)}_{\mu\nu} = V_{QQg}^\sigma G_Q 
\gamma_\mu P_L K_{\sigma\nu}.
\end{equation}

Once again we separate out the temporal and spatial contributions to the 
diagram, akin to Eq.~\eqref{eq:ac0}, and determine the mixing coefficients 
from
\begin{equation}
c_{11}^{\mathrm{latt}} = c_k,\quad \mathrm{and}\quad c_{12}^{\mathrm{latt}} = 
c_k - c_0.
\end{equation}

The procedure for diagram (a$^\prime$) is much the same. This time the starting 
point is (note the different spinor structure)
\begin{equation}
(\mathrm{a}^\prime) = \frac{1}{2}\delta_{AB}\delta_{CD} 
\Big(\overline{u}_Q{\cal 
V}^{(1)\,\mu\nu} v_q\Big)\Big(\overline{v}_{\overline{Q}}{\cal 
V}^{(2)}_{\mu\nu}  u_q\Big),
\end{equation}
with ${\cal V}^{(1)\,\mu\nu}$ and ${\cal V}^{(2)}_{\mu\nu}$ given in Eq.~
\eqref{eq:vmunu}.

For these diagrams, we confirmed that the contributions from 
the pairs of diagrams (c) and (d), (a$^\prime$) and (b$^\prime$), and 
(c$^\prime$) 
and (d$^\prime$), are each identical.

\subsubsection{Operators $\widehat{Q}3$ and $\widehat{Q}5$}

The previous discussion focused on the extraction of the mixing 
coefficients for $\widehat{Q}1$, $\widehat{Q}2$, $\widehat{Q}4$, 
and $\widehat{Q}6$, which all have the same color structure. The contributions 
from $\widehat{Q}3$, $\widehat{Q}5$ and $\widehat{Q}7$ have a different color 
structure. It is straightforward to the observer, however, that by judicious 
choice 
of external colors and appropriate Fierz relations, the contributions to these 
operators can be related to those from operators $\widehat{Q}2$, 
$\widehat{Q}4$, and $\widehat{Q}6$.

For example, one can compare the term proportional to $\delta_{AB}\delta_{CD}$ 
for $\widehat{Q}2$ with that proportional to $\delta_{AD}\delta_{BC}$ 
for $\widehat{Q}3$ and then, taking into account the relative color factors, 
one finds
\begin{align}
c_{33}^{\mathrm{latt},\,(a)/(b)} = {} & 
\frac{1}{3}c_{33}^{\mathrm{latt},\,(c^\prime)/(d^\prime)} =  
-\frac{1}{8}c_{22}^{\mathrm{latt},\,(a)/(b)}, \nonumber\\
c_{33}^{\mathrm{latt},\,(c)/(d)} = {} & 
-8c_{22}^{\mathrm{latt},\,(c)/(d)},\nonumber\\
c_{33}^{\mathrm{latt},\,(e)} ={} 
&  c_{22}^{\mathrm{latt},(e)}, \qquad \;
c_{33}^{\mathrm{latt},\,(e^\prime)} =  
c_{22}^{\mathrm{latt},\,(e^\prime)},\nonumber\\
c_{33}^{\mathrm{latt},\,(f)} =  {} &
c_{22}^{\mathrm{latt},\,(f)},\qquad   c_{33}^{\mathrm{latt},\,(f^\prime)} =  
c_{22}^{\mathrm{latt},\,(f^\prime)}.
\end{align}

We have verified by explicit calculation for a specific choice of heavy 
quark mass that these relations hold.

Combined with the appropriate Fierz identities, these results 
reduce the number of integrations we must carry out. This 
significantly speeds up the matching procedure, because there are 
approximately 80 nonzero coefficient contributions that must be determined at 
each heavy quark mass for the complete matching calculation.

\subsection{Dimension seven operators}

\begin{figure}
\includegraphics[width=0.45\textwidth,keepaspectratio=true]{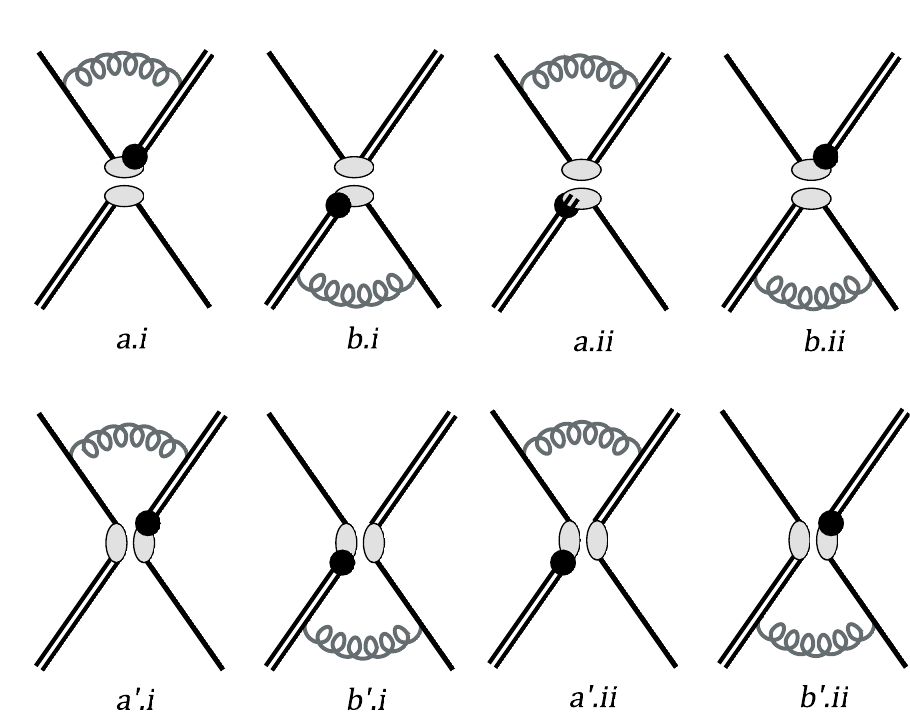}
\caption{\label{fig:oneloopj1} 
Sample one loop diagrams representing the corrections to matrix 
elements of 
the $1/M$ operators $\widehat{Q}j1$. The black dot represents a derivative 
acting on the heavy (anti)quark propagator. The external states are those of 
Fig.~\protect\ref{fig:tree}. We show the corrections associated with diagrams 
(a), (b), 
(a$^\prime$), and (b$^\prime$) of 
Fig.~\protect\ref{fig:oneloop}. Analogous diagrams exist for diagrams (c) to 
f$^\prime$). In general diagrams such as a$.ii$ and b$.ii$ vanish, because the 
derivative acts on an external heavy (anti-) quark with zero momentum.}
\end{figure}

We represent the diagrams that include the $1/M$ operators, $\widehat{Q}i1$, in 
Fig.~\ref{fig:oneloopj1}. Note that diagrams in which 
the derivative acts directly on an external heavy quark or antiquark vanish, 
because these external states have zero spatial momentum.

\begin{figure}
\includegraphics[width=0.45\textwidth,keepaspectratio=true]{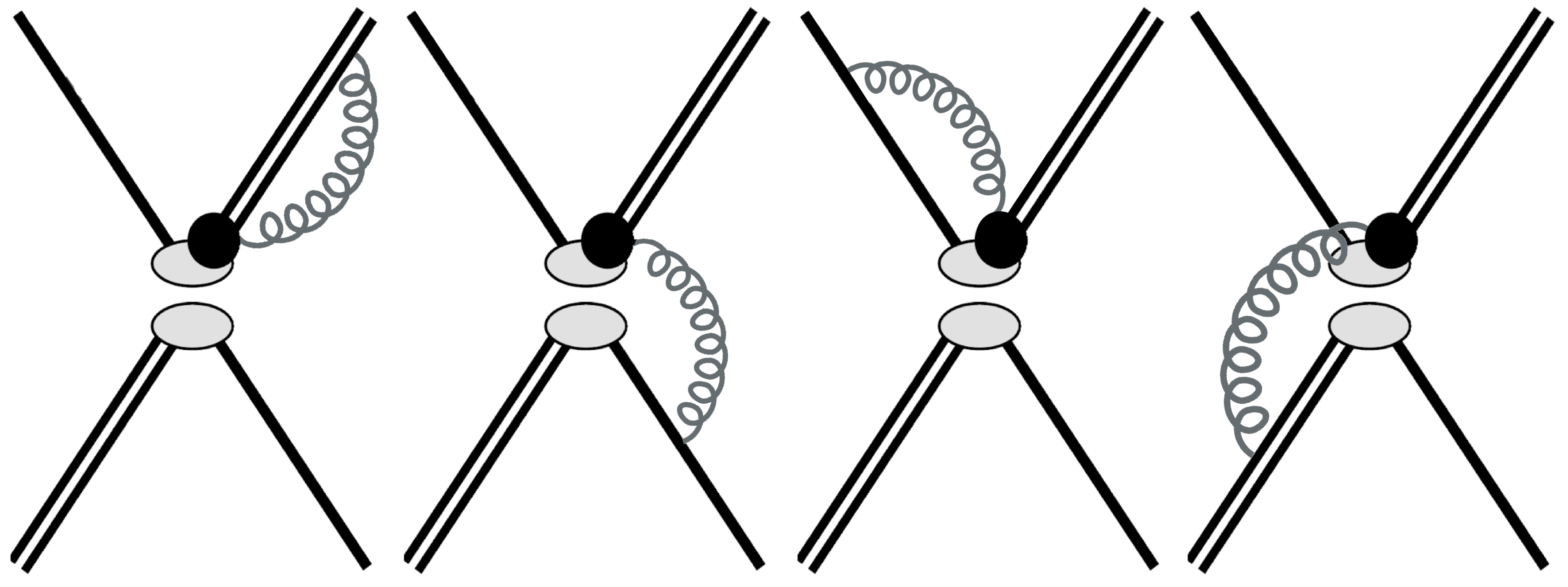}
\caption{\label{fig:alpham} 
Sample one loop diagrams representing the one loop corrections to matrix 
elements of 
the $1/M$ operators $\widehat{Q}j1$. We show the four corrections associated 
with diagram (a) of Fig.~\protect\ref{fig:oneloop}. Analogous diagrams exist 
for 
diagrams (b) to (f$^\prime$). For more details, see the caption of 
Fig.~\protect\ref{fig:oneloop}. We do not include these contributions in our 
matching procedure.}
\end{figure}

We expect that the systematic truncation uncertainty is 
dominated by missing terms of ${\cal O}(\alpha_s^2)$ and therefore we do not 
include contributions that appear at ${\cal 
O}(\alpha_s\Lambda_{\mathrm{QCD}}/M_b)$, which we illustrate in Fig.~
\ref{fig:alpham}. These contributions are generated by gluon emission at the 
$1/M$ operator vertex and, to our knowledge, have not been calculated in 
continuum QCD.

The extraction of the mixing coefficients, $\zeta_{ij}^{\mathrm{latt}}$, for 
the $1/M$ operators parallels that for the leading order operators, with 
two small differences. The first is the inclusion of a derivative acting on the 
heavy (anti-)quark propagator. The second is the presence of the extra gamma 
matrix in the operator, which means that the result is extracted from the 
coefficient of a different element of the Dirac algebra than in the leading 
order case. These changes aside, the process is the same.

The results are all infrared finite, which we confirm by explicit calculation 
at different values of the gluon masses. Furthermore we verify that 
\begin{equation}
\zeta_{11}^{\mathrm{latt},\,(a)/(b)}=\zeta_{10}^{(V_k)},\qquad\mathrm{and}
\qquad 
\zeta_{12}^{\mathrm{latt},\,(a)/(b)}=\zeta_{10}^{(V_k)}-\zeta_{10}^{(V_0)},
\end{equation}
where the $\zeta_{10}^{(V_\mu)}$ are the heavy-light current 
matching results of \cite{Monahan:2012dq}.

\subsection{Wavefunction renormalization}

To complete the matching calculation we also require the HISQ 
and NRQCD wavefunction renormalization contributions. The one loop parameters 
of 
NRQCD have 
been extensively studied in the
literature, for example in 
\cite{morningstar93,Dalgic:2003uf,Dowdall:2011wh,Monahan:2012dq} and we 
describe the complete one loop calculations for both massless and massive HISQ 
in \cite{Monahan:2012dq}. 
For the purposes of this paper, we need only the massless HISQ result:
\begin{equation}
Z_q = 1-\als\!\left[C_q+ 
\frac{1}{3\pi}\left[1-\left(1-\xi\right)\right]
\log\left(a^2\lambda^2\right)\right]\! +\!{\cal O}(\als^2),
\end{equation}
where $a\lambda$ is a gluon mass introduced to regulate the infrared 
divergence. Here $\xi$ is the gauge-fixing parameter: for Feynman gauge, $\xi 
=1$. The infrared finite contribution, $C_q$, is $C_q =  0.3940(3)$ in Feynman 
gauge.

The NRQCD 
wavefunction renormalization, $Z_H$, is given by
\begin{equation}
Z_H = 1 + \als\!\left[C_H-\frac{1}{3\pi}\left[2+\left(1-\xi\right)\right]
\log\left(a^2\lambda^2\right)\right] + {\cal
O}(\als^2).
\end{equation}

We tabulate the infrared
finite contribution, $C_H$, in Table
\ref{tab:nrqcdparms}. We present results with the tree-level NRQCD 
coefficients, $c_i = 1$, and use the Landau
link definition of the tadpole improvement factor $u_0$, with
$u_0^{(1)}=0.7503(1)$. All results use
stability parameter $n=4$.

\begin{table}
\caption{\label{tab:nrqcdparms}Infrared finite contributions to the one loop 
wavefunction renormalization in NRQCD.
All results use stability parameter $n=4$. We implement tadpole improvement
with the Landau link definition of $u_0$. All results are in Feynman gauge.
The statistical
uncertainties from the numerical integration of the relevant diagrams are
unity in the final digit. \\}
\begin{ruledtabular}
\begin{tabular}{cccccccc}
\vspace*{-5pt}\\
$aM_0$ & 3.297 & 3.263 & 3.25\hphantom{0} & 2.66\hphantom{0} & 2.62\hphantom{0} 
& 1.91\hphantom{0} & 1.89\hphantom{0} \\
\vspace*{-5pt}\\
\hline
\vspace*{-5pt}\\
$C_H$ & -0.235 & -0.241 & -0.244 & -0.366 & -0.374 & -0.617 & -0.627 \\
\vspace*{-5pt}\\
\end{tabular}
\end{ruledtabular}
\end{table}

In the following, we incorporate the wavefunction renormalizations, $Z_q$ and 
$Z_Q$, in the mixing coefficients $c_{ij}$ with $i=j$.

\section{\label{sec:matchresults}Results}

\subsection{\label{sec:contqcdres}In continuum QCD}

The mixing coefficients defined in Eq.~\eqref{eq:qcdcij}, $c_{ij}$, are 
given to ${\cal O}(\alpha_s)$ in \cite{Gamiz:2008sk}. Coefficients $c_{11}$, 
$c_{12}$, $c_{22}$, and $c_{21}$ were first published in 
\cite{Hashimoto:2000eh}. Here we collect the results for the mixing 
coefficients for completeness. We discuss the continuum one loop 
calculation in more detail in the Appendix, where we focus 
on the scheme dependence of the ``evanescent'' operators that enter the 
matching procedure and correct Eqs.~(B9) and (B10) of \cite{Gamiz:2008sk}.

The nonzero coefficients for the standard model 
operators in the ``BBGLN'' scheme of \cite{Beneke:1998sy} are
\begin{align}
c_{11} = {} & 
\frac{1}{4\pi}\left\{-\frac{35}{3}-2\log\frac{\mu^2}{M^2} 
-4\log\frac{\lambda^2}{ M^2}\right\}, \label{eq:c11} \\
c_{12} = {} & -\frac{8}{4\pi}, \label{eq:c12} \\
c_{22} = {} & 
\frac{1}{4\pi}\left\{10+\frac{16}{3}\log\frac{\mu^2}{M^2} 
-\frac{4}{3}\log\frac{\lambda^2}{ M^2}\right\}, \label{eq:c22} \\
c_{21} = {} & 
\frac{1}{4\pi}\left\{\frac{3}{2}+\frac{1}{3}\log\frac{\mu^2}{M^2} 
+\frac{2}{3}\log\frac{\lambda^2}{ M^2}\right\}, \label{eq:c21} \\
c_{33} = {} & 
\frac{1}{4\pi}\left\{-2-\frac{8}{3}\log\frac{\mu^2}{M^2} 
-\frac{4}{3}\log\frac{\lambda^2}{ M^2}\right\}, \label{eq:c33} \\
c_{31} = {} & 
\frac{1}{4\pi}\left\{3+\frac{4}{3}\log\frac{\mu^2}{M^2} 
+\frac{2}{3}\log\frac{\lambda^2}{ M^2}\right\}, \label{eq:c31}
\end{align}
while the mixing coefficients for the remaining operators in the SUSY basis are
\begin{align}
c_{44} = {} & 
\frac{1}{4\pi}\left\{\frac{143}{12}+8\log\frac{\mu^2}{M^2} 
-\frac{7}{2}\log\frac{\lambda^2}{ M^2}\right\}, \label{eq:c44} \\
c_{45} = {} & 
\frac{1}{4\pi}\left\{-\frac{23}{4}-\frac{3}{2}\log\frac{\lambda^2}{ 
M^2}\right\}, \label{eq:c45} \\
c_{55} = {} & 
\frac{1}{4\pi}\left\{-\frac{85}{12}-\log\frac{\mu^2}{M^2} 
-\frac{7}{2}\log\frac{\lambda^2}{ M^2}\right\}, \label{eq:c55} \\
c_{54} = {} & 
\frac{1}{4\pi}\left\{\frac{13}{4}+3\log\frac{\mu^2}{M^2} 
-\frac{3}{2}\log\frac{\lambda^2}{ M^2}\right\}. \label{eq:c54}
\end{align}

In addition, we require the mixing coefficients for the intermediate operators 
$Q6$ and $Q7$ of Eqs.~\eqref{eq:q6} and \eqref{eq:q7}, which are given by
\begin{align}
c_{46} = {} & 
\frac{1}{4\pi}\left\{\frac{23}{8}+\frac{3}{4}\log\frac{\lambda^2}{ 
M^2}\right\}. 
\label{eq:c46} \\
c_{57} = {} & 
\frac{1}{4\pi}\left\{-\frac{13}{8}-\frac{3}{2}\log\frac{\mu^2}{M^2} 
+\frac{3}{4}\log\frac{\lambda^2}{ M^2}\right\}.\label{eq:c57} 
\end{align}

\subsection{On the lattice}

\begin{figure}
\includegraphics[width=0.52\textwidth,keepaspectratio=true]{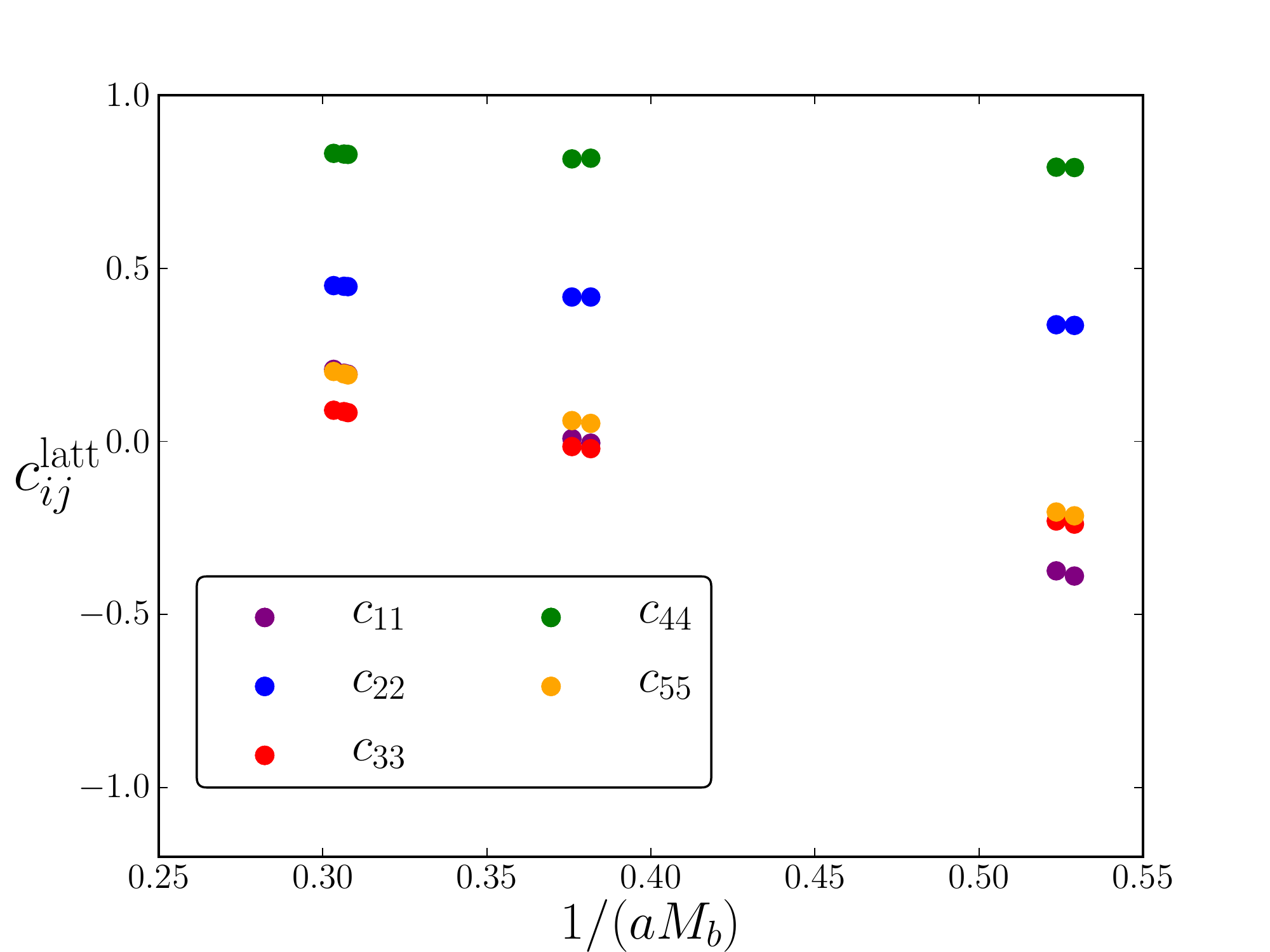}
\caption{\label{fig:cii} 
Mass dependence of the lattice coefficients $c_{ij}^{\mathrm{latt}}$, for 
$i=j$ [\emph{color online}]. Statistical uncertainties from 
numerical integration are $\pm$0.002 and smaller than the data points on this 
scale.}
\end{figure}

\begin{figure}
\includegraphics[width=0.52\textwidth,keepaspectratio=true]{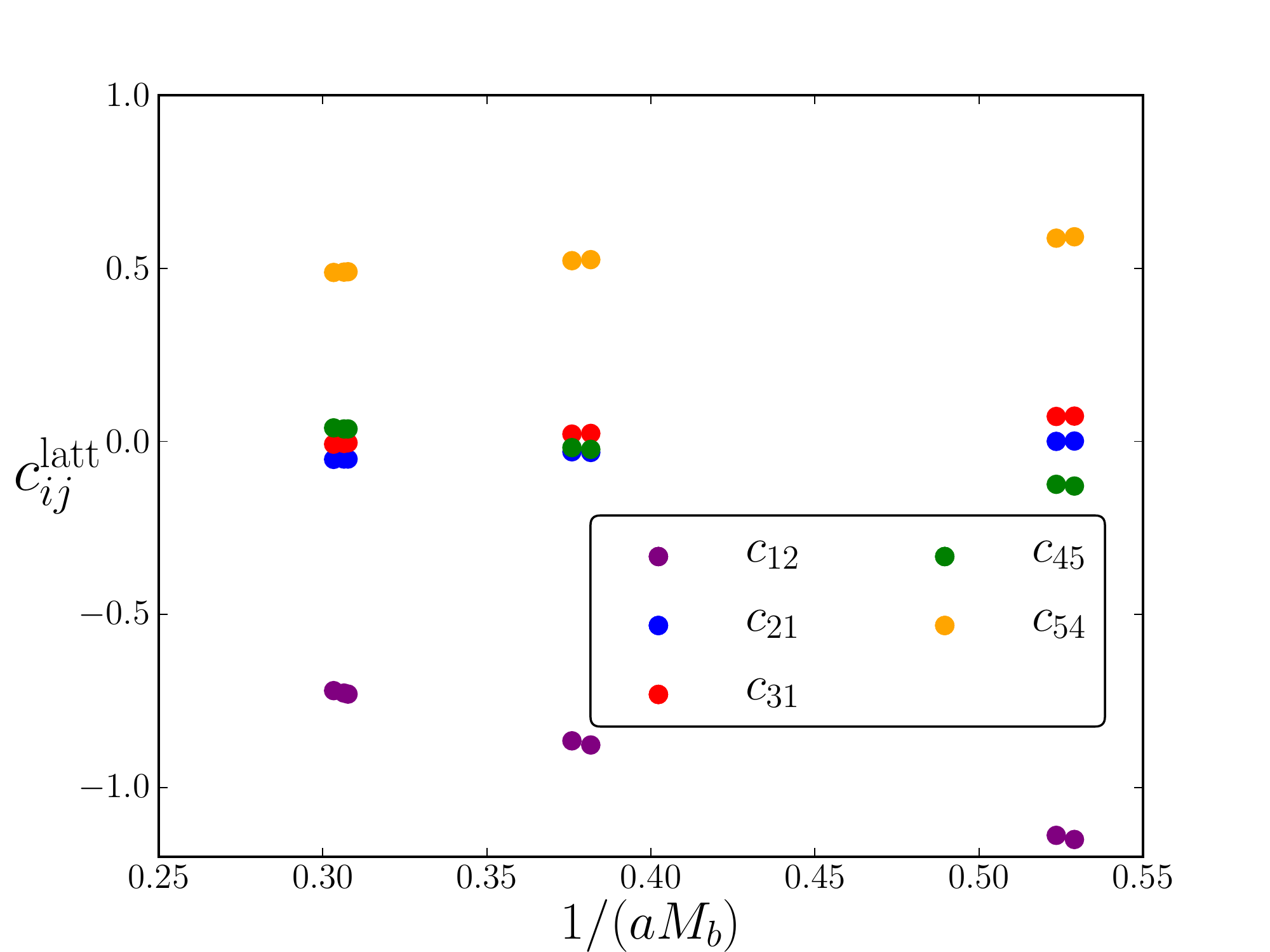}
\caption{\label{fig:cij} 
Mass dependence of the lattice coefficients $c_{ij}^{\mathrm{latt}}$, for 
$i\neq j$ [\emph{color online}]. Statistical 
uncertainties from 
numerical integration are $\pm$0.002 and smaller than the data points on this 
scale.}
\end{figure}

We tabulate the infrared finite contributions to the one loop lattice 
coefficients in Table \ref{tab:cijlatt}. 
For a breakdown of the individual contributions to the mixing 
coefficients, see \cite{Gamiz:2008sk}, which demonstrates how one 
obtains the final result for $c_{44}^{\mathrm{latt}}$ and 
 $c_{46}^{\mathrm{latt}}$ and recovers the continuum infrared behavior.
\begin{table}
\caption{\label{tab:cijlatt}One loop lattice coefficients, 
$c_{ij}^{\mathrm{latt}}$, for HISQ-NRQCD $\Delta B=2$ operators. We include 
only 
the infrared finite contributions to the coefficients.
The statistical
uncertainties from the numerical integration of the relevant diagrams are
$\pm$0.002. \\}
\begin{ruledtabular}
\begin{tabular}{cccccccc}
\vspace*{-5pt}\\
$aM_0$ & 3.297 & 3.263 & 3.25\hphantom{0} & 2.66\hphantom{0} & 2.62\hphantom{0} 
& 1.91\hphantom{0} & 1.89\hphantom{0} \\
\vspace*{-5pt}\\
\hline
\vspace*{-5pt}\\
$c_{11}^{\mathrm{latt}}$ & \hphantom{-}0.208 & \hphantom{-}0.197 & 
\hphantom{-}0.194 & \hphantom{-}0.008 & -0.005 & -0.374 & 
-0.389 \\
$c_{12}^{\mathrm{latt}}$ & -0.720 & -0.727 & -0.730 & -0.865 & -0.877 & -1.138 
& 
-1.150 \\
$c_{22}^{\mathrm{latt}}$ & \hphantom{-}0.450 & \hphantom{-}0.448 & 
\hphantom{-}0.447 &  \hphantom{-}0.417 & \hphantom{-}0.417 & \hphantom{-}0.337
& 
\hphantom{-}0.335 \\
$c_{21}^{\mathrm{latt}}$ & -0.052 & -0.051 & -0.051 & -0.030 & -0.032 & 
\hphantom{-}0.000 & \hphantom{-}0.001 \\
$c_{33}^{\mathrm{latt}}$ & \hphantom{-}0.090 & \hphantom{-}0.086 & 
\hphantom{-}0.083 & -0.015 & -0.021 & -0.230 & 
-0.239 \\
$c_{31}^{\mathrm{latt}}$ & -0.008 & -0.006 & -0.004 & \hphantom{-}0.021 & 
\hphantom{-}0.023 & \hphantom{-}0.072 & \hphantom{-}0.073 \\
$c_{44}^{\mathrm{latt}}$ & \hphantom{-}0.832 & \hphantom{-}0.830 & 
\hphantom{-}0.829 & \hphantom{-}0.816 & \hphantom{-}0.818 & \hphantom{-}0.792 & 
\hphantom{-}0.791  \\
$c_{45}^{\mathrm{latt}}$ & \hphantom{-}0.039 & \hphantom{-}0.036 & 
\hphantom{-}0.036 &  -0.018 & -0.023 & -0.124 & -0.129 \\
$c_{55}^{\mathrm{latt}}$ & \hphantom{-}0.202 & \hphantom{-}0.195 & 
\hphantom{-}0.192 &  \hphantom{-}0.060 & \hphantom{-}0.052 & -0.204 
& 
-0.215 \\
$c_{54}^{\mathrm{latt}}$ & \hphantom{-}0.488 & \hphantom{-}0.489 & 
\hphantom{-}0.490 & \hphantom{-}0.522 & \hphantom{-}0.525 & \hphantom{-}0.587 & 
\hphantom{-}0.591 \\
\vspace*{-5pt}\\
\end{tabular}
\end{ruledtabular}
\end{table}
For illustration, we plot the mass dependence of the coefficients 
$c_{ij}^{\mathrm{latt}}$, for $i=j$ and $i \neq j$, in Figs.~\ref{fig:cii} 
and \ref{fig:cij}, respectively. Note that the scales on the vertical axes of 
these two plots are identical, to enable easy comparison.

\subsection{Matching coefficients}

\begin{figure}
\includegraphics[width=0.5\textwidth,keepaspectratio=true]{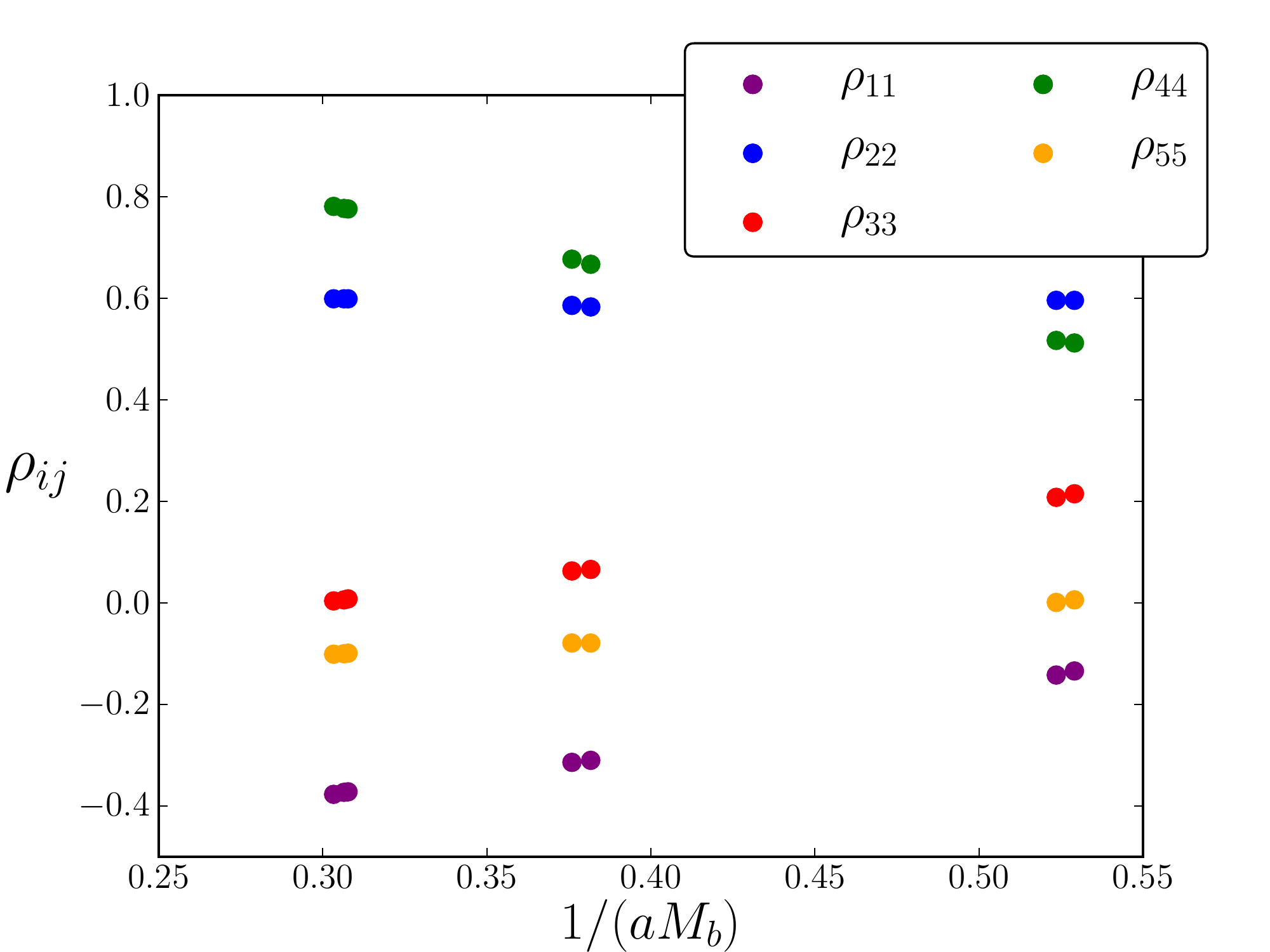}
\caption{\label{fig:rhoii} 
Mass dependence of the leading order matching coefficients $\rho_{ij}$, for 
$i=j$ [\emph{color online}].  Statistical 
uncertainties from 
numerical integration are $\pm$0.002 and smaller than the data points on this 
scale.}
\end{figure}

\begin{figure}
\includegraphics[width=0.52\textwidth,keepaspectratio=true]{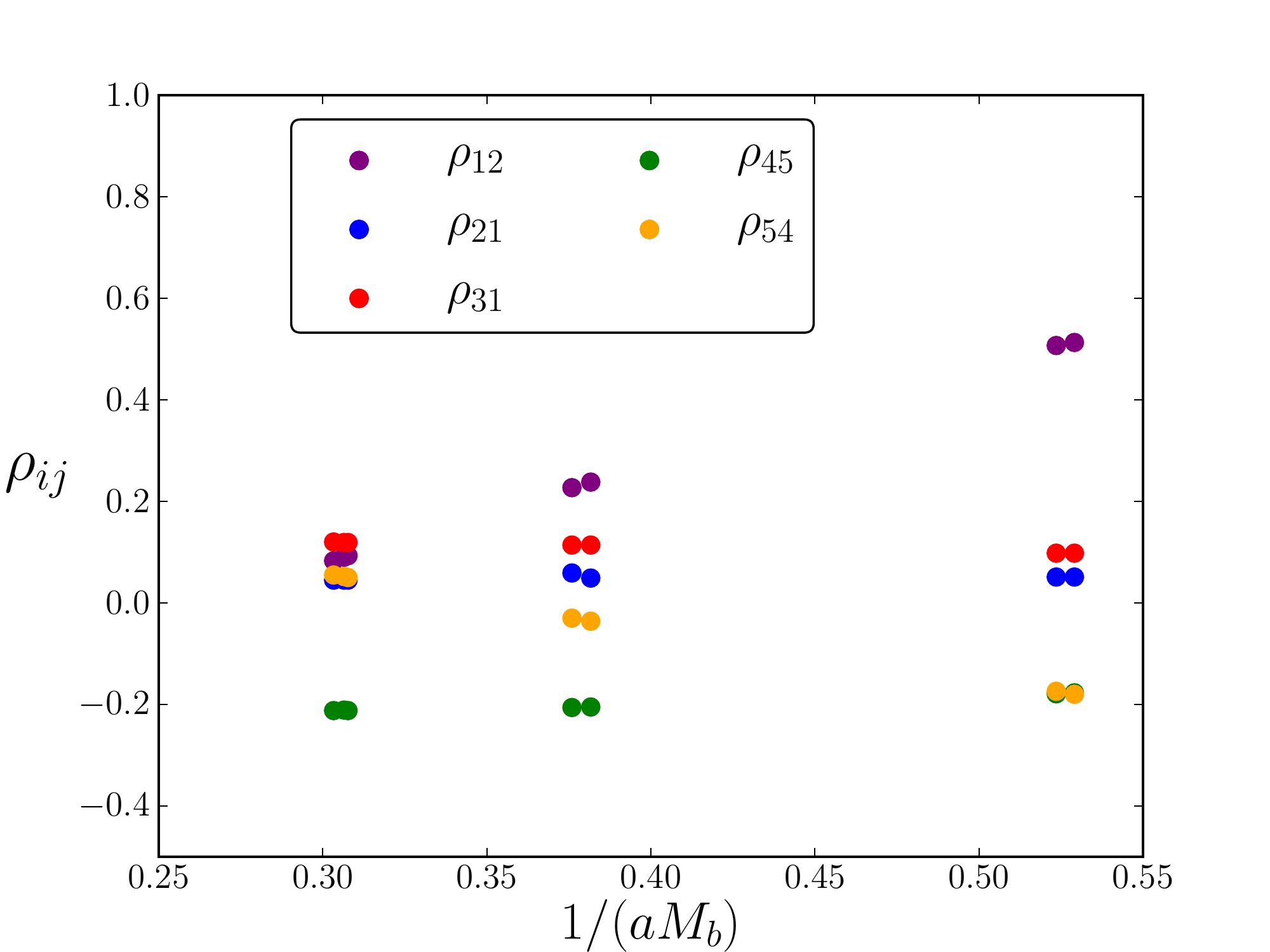}
\caption{\label{fig:rhoij} 
Mass dependence of the matching coefficients $\rho_{ij}$, for 
$i\neq j$ [\emph{color online}].}
\end{figure}

\begin{figure}
\includegraphics[width=0.52\textwidth,keepaspectratio=true]{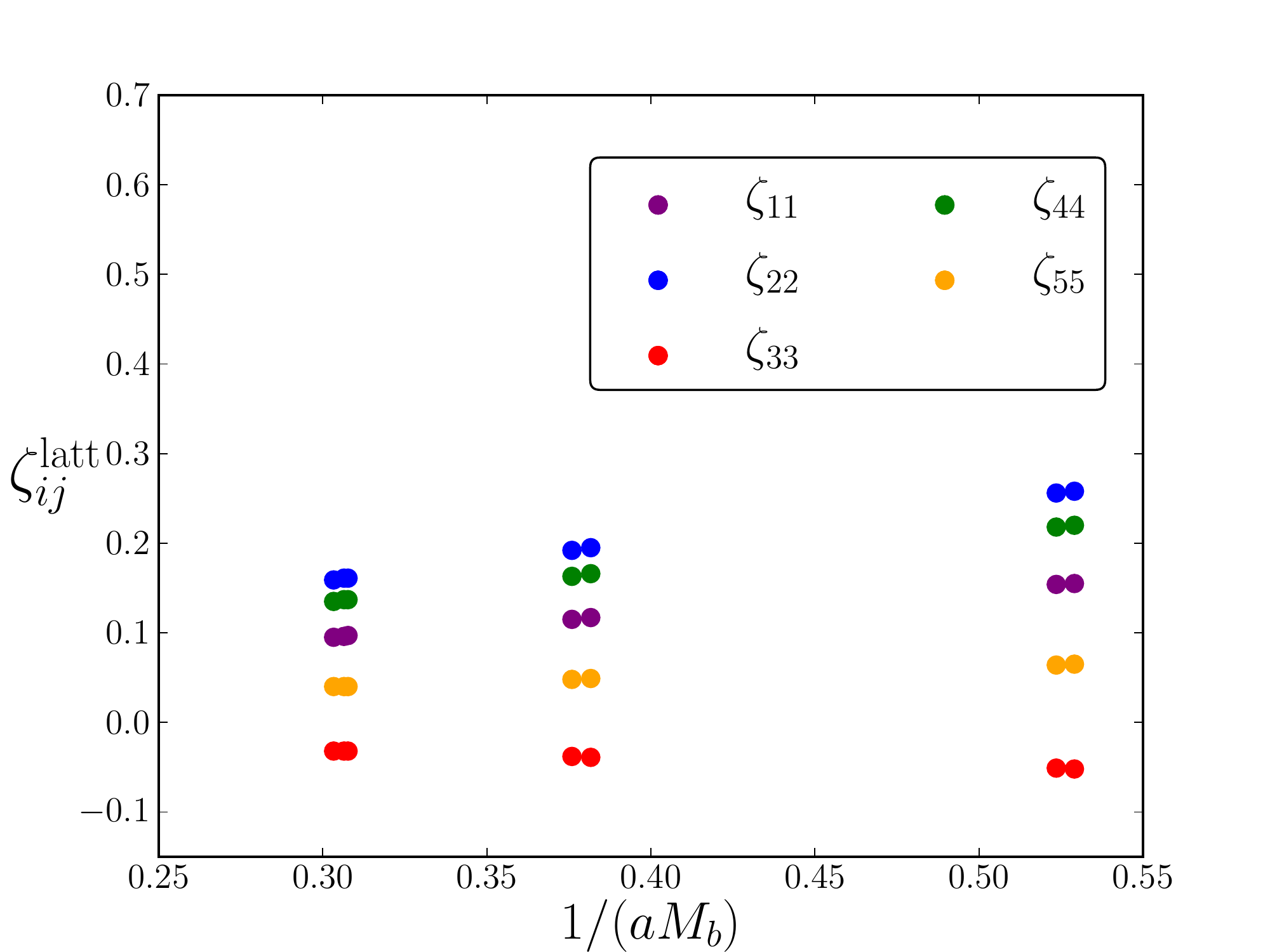}
\caption{\label{fig:zetaii} 
Mass dependence of the next-to-leading order matching coefficients 
$\zeta_{ij}$, for 
$i=j$ [\emph{color 
online}].  Statistical 
uncertainties from 
numerical integration are $\pm$0.002 and smaller than the data points on this 
scale.}
\end{figure}

\begin{figure}
\includegraphics[width=0.52\textwidth,keepaspectratio=true]{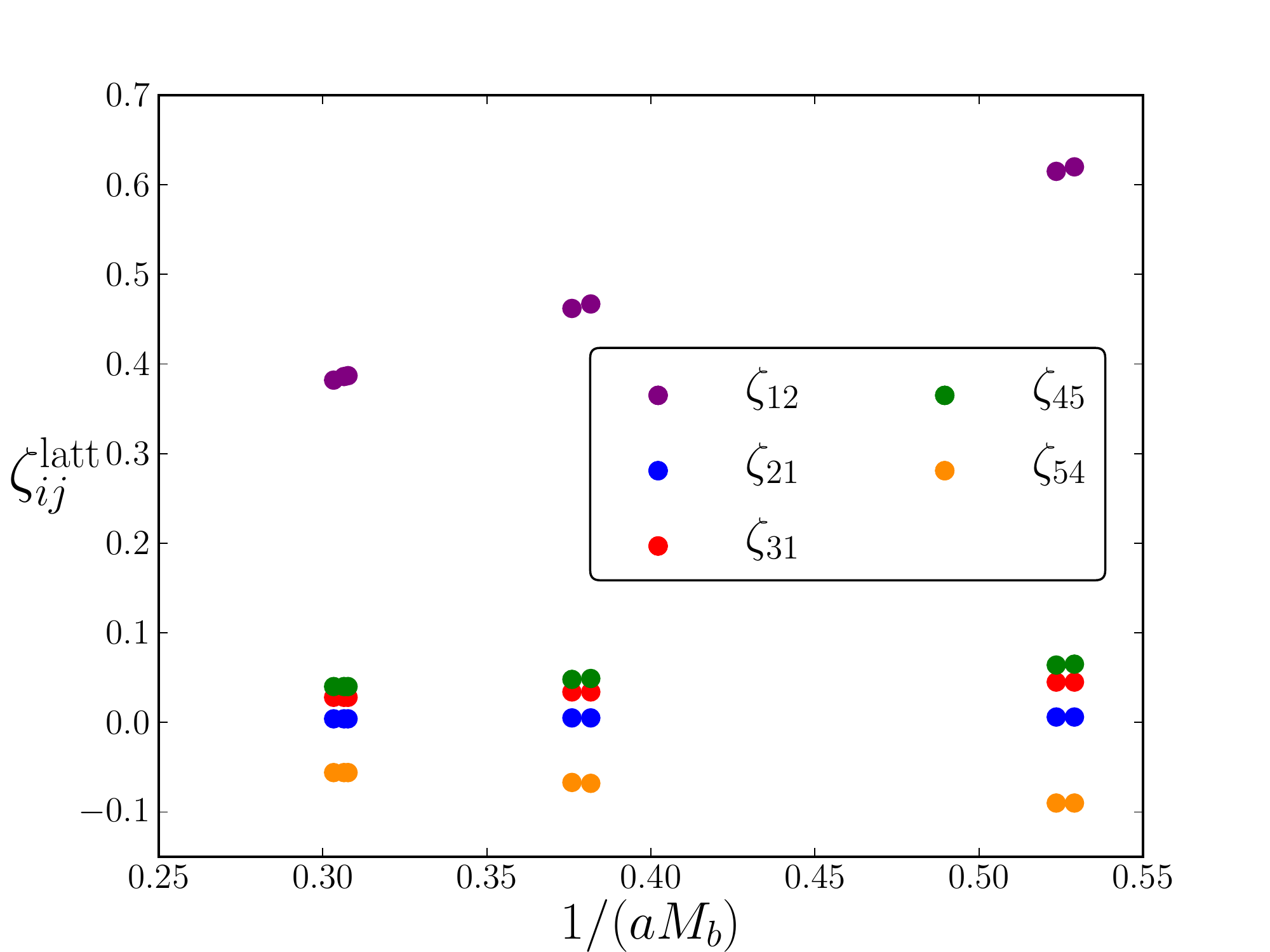}
\caption{\label{fig:zetaij} 
Mass dependence of the next-to-leading order matching coefficients 
$\zeta_{ij}$, 
for 
$i\neq j$ [\emph{color 
online}]. Statistical 
uncertainties from 
numerical integration are $\pm$0.002 and smaller than the data points on this 
scale.}
\end{figure}

Tables \ref{tab:rhoij} and \ref{tab:zetaij} summarize the final results of our 
calculation. Table \ref{tab:rhoij} lists the leading-order matching 
coefficients, $\rho_{ij}$, at a range of heavy quark masses and at a scale 
equal to the heavy quark mass. We tabulate
next-to-leading contributions, $\zeta_{ij}$, in Table \ref{tab:zetaij}. We plot 
the leading order coefficients $\rho_{ij}$ in Figs.~
\ref{fig:rhoii} and \ref{fig:rhoij}, and the 
next-to-leading order coefficients $\zeta_{ij}$ in Figs.~
\ref{fig:zetaii} and \ref{fig:zetaij}. We use the same vertical axes 
to simplify comparison between Figs.~
\ref{fig:rhoii} and \ref{fig:rhoij} and between Figs.~
\ref{fig:zetaii} and \ref{fig:zetaij}. We choose heavy quark masses that 
correspond to the HPQCD collaboration's ongoing nonperturbative calculations of 
neutral $B$ mixing \cite{inprep}. These masses are a subset of those presented 
in the matching calculation of \cite{Monahan:2012dq}.

\begin{table}
\caption{\label{tab:rhoij}One loop matching coefficients for HISQ-NRQCD 
$\Delta B=2$ operators.
The statistical
uncertainties from the numerical integration of the relevant diagrams are
$\pm$0.002. \\}
\begin{ruledtabular}
\begin{tabular}{cccccccc}
\vspace*{-5pt}\\
$aM_0$ & 3.297 & 3.263 & 3.25\hphantom{0} & 2.66\hphantom{0} & 2.62\hphantom{0} 
& 1.91\hphantom{0} & 1.89\hphantom{0} \\
\vspace*{-5pt}\\
\hline
\vspace*{-5pt}\\
$\rho_{11}$ & -0.377 & -0.373 & -0.372 & -0.314 & -0.310 & -0.142 & -0.134 \\
$\rho_{12}$ & \hphantom{-}0.083 & \hphantom{-}0.090 & \hphantom{-}0.093 & 
\hphantom{-}0.227 & \hphantom{-}0.238 & \hphantom{-}0.507 & \hphantom{-}0.513 \\
$\rho_{22}$ & \hphantom{-}0.599 & \hphantom{-}0.599 & \hphantom{-}0.599 &  
\hphantom{-}0.586 & \hphantom{-}0.583 & \hphantom{-}0.596 & \hphantom{-}0.596 \\
$\rho_{21}$ & \hphantom{-}0.045 & \hphantom{-}0.045 & \hphantom{-}0.045 & 
\hphantom{-}0.059 & \hphantom{-}0.049 & \hphantom{-}0.051 & 
\hphantom{-}0.051 \\
$\rho_{33}$ & \hphantom{-}0.004 & \hphantom{-}0.006 & 
\hphantom{-}0.008 & \hphantom{-}0.063 & \hphantom{-}0.066 & \hphantom{-}0.208 & 
\hphantom{-}0.215 \\
$\rho_{31}$ & \hphantom{-}0.120 & \hphantom{-}0.119 & \hphantom{-}0.119 & 
\hphantom{-}0.114 & \hphantom{-}0.114 & \hphantom{-}0.098 & \hphantom{-}0.098 \\
$\rho_{44}$ & \hphantom{-}0.781 & \hphantom{-}0.777 & \hphantom{-}0.776 & 
\hphantom{-}0.677 & \hphantom{-}0.667 & \hphantom{-}0.517 & \hphantom{-}0.512 \\
$\rho_{45}$ & -0.212 & -0.211 & -0.212 & -0.206 & -0.205 & -0.179 & -0.177 \\
$\rho_{55}$ & -0.101 & -0.100 & -0.099 &  -0.079 & -0.079 & \hphantom{-}0.001 & 
\hphantom{-}0.006 \\
$\rho_{54}$ & \hphantom{-}0.055 & \hphantom{-}0.052 & \hphantom{-}0.050 & 
-0.030 & -0.036 & -0.174 & -0.180 \\
\vspace*{-5pt}\\
\end{tabular}
\end{ruledtabular}
\end{table}

\begin{table}
\caption{\label{tab:zetaij}Next-to-leading order matching coefficients for 
HISQ-NRQCD $\Delta B=2$ operators.
The statistical
uncertainties from the numerical integration of the relevant diagrams are
$\pm$0.002. \\}
\begin{ruledtabular}
\begin{tabular}{cccccccc}
\vspace*{-5pt}\\
$aM_0$ & 3.297 & 3.263 & 3.25\hphantom{0} & 2.66\hphantom{0} & 2.62\hphantom{0} 
& 1.91\hphantom{0} & 1.89\hphantom{0} \\
\vspace*{-5pt}\\
\hline
\vspace*{-5pt}\\
$\zeta_{11}$ & \hphantom{-}0.095 & \hphantom{-}0.096 & \hphantom{-}0.097 & 
\hphantom{-}0.115 & \hphantom{-}0.117 & \hphantom{-}0.154 & \hphantom{-}0.155 \\
$\zeta_{12}$ & \hphantom{-}0.382 & \hphantom{-}0.386 & \hphantom{-}0.387 & 
\hphantom{-}0.462 & \hphantom{-}0.467 & \hphantom{-}0.615 & \hphantom{-}0.620 \\
$\zeta_{22}$ & \hphantom{-}0.159 & \hphantom{-}0.161 & \hphantom{-}0.161 &  
\hphantom{-}0.192 & \hphantom{-}0.165 & \hphantom{-}0.256 & \hphantom{-}0.258 \\
$\zeta_{21}$ & \hphantom{-}0.004 & \hphantom{-}0.004 & \hphantom{-}0.004 & 
\hphantom{-}0.005 & \hphantom{-}0.005 & \hphantom{-}0.006 & \hphantom{-}0.006 \\
$\zeta_{33}$ & -0.032 & -0.032 & -0.032 & -0.038 & -0.039 & -0.051 & -0.052 \\
$\zeta_{31}$ & \hphantom{-}0.028 & \hphantom{-}0.028 & \hphantom{-}0.028 & 
\hphantom{-}0.034 & \hphantom{-}0.034 & \hphantom{-}0.045 & \hphantom{-}0.045 \\
$\zeta_{44}$ & \hphantom{-}0.135 & \hphantom{-}0.137 & \hphantom{-}0.137 & 
\hphantom{-}0.163 & \hphantom{-}0.166 & \hphantom{-}0.218 & \hphantom{-}0.220 \\
$\zeta_{45}$ & \hphantom{-}0.040 & \hphantom{-}0.040 & \hphantom{-}0.040 &  
\hphantom{-}0.048 & \hphantom{-}0.049 & \hphantom{-}0.064 & \hphantom{-}0.065 \\
$\zeta_{55}$ & \hphantom{-}0.040 & \hphantom{-}0.040 & \hphantom{-}0.040 &  
\hphantom{-}0.048 & \hphantom{-}0.049 & \hphantom{-}0.064 & \hphantom{-}0.065 \\
$\zeta_{54}$ & -0.056 & -0.056 & -0.056 & -0.067 & -0.068 & -0.090 & -0.090 \\
\vspace*{-5pt}\\
\end{tabular}
\end{ruledtabular}
\end{table}

\section{\label{sec:summary}Summary}

We have determined the one loop matching coefficients required to match 
the matrix elements of heavy-light four-fermion operators on the lattice to 
those in continuum QCD. We used NRQCD for the 
heavy quarks and massless HISQ light quarks. We incorporated the full set of 
five independent $\Delta B=2$ operators 
relevant to neutral $B$ mixing both within and beyond the standard model and 
carried out the matching procedure through ${\cal 
O}(\alpha_s,\Lambda_{\mathrm{QCD}}/M_b,\alpha_s/(aM_b))$. The 
perturbative coefficients are well behaved and all are smaller than unity. 

The dominant systematic uncertainties in our matching procedure appear at 
${\cal O}(\alpha_s^2)$ with next-to-leading contributions at ${\cal 
O}(\alpha_s\Lambda_{\mathrm{QCD}}/M_b)$, the exact values of which will depend 
on the choice of lattice spacing and matching scale. We estimate 
that these uncertainties will correspond to a systematic uncertainty of 
approximately a few percent in 
the final results for nonperturbative matrix elements in the $\overline{MS}$ 
scheme \cite{Gamiz:2009ku}. We note that the uncertainties arising from 
perturbative matching will be significantly reduced in ratios of 
nonperturbative matrix elements \cite{Gamiz:2009ku,Na:2012kp} and that, in 
general, many HISQ parameters exhibit better perturbative convergence than 
their 
AsqTad counterparts \cite{Monahan:2012dq}.

These matching coefficients are critical ingredients in the determination
of neutral $B$ meson mixing on the lattice using NRQCD and HISQ 
quarks. Without these coefficients, 
matrix elements calculated nonperturbatively on the lattice cannot be 
related to experimentally relevant results in continuum QCD in the 
$\overline{MS}$ scheme. Since any lattice calculation of neutral meson 
mixing that incorporates an effective theory description of the heavy quark 
requires some matching procedure, we have included full details of the lattice 
perturbation theory calculation, not previously available in the 
literature, as an aid to future calculations.

Although recent work on the decays of the $B_s$ meson 
has been carried out using the relativistic HISQ action for $b$ and $s$ quarks 
\cite{McNeile:2011ng}, such calculations are currently prohibitively expensive 
for the $B_d$ meson. Furthermore, computations at the physical $b$ quark mass 
are not yet possible and an HQET-guided expansion up to the physical point is 
still required. Therefore, the use of effective theories for heavy-light 
systems remains the most practical method for precise predictions of neutral 
$B$ meson mixing phenomena.

% Use the proper section head for acknowledgments.
\begin{acknowledgments}
The authors are particularly grateful to Christine Davies for useful 
discussions regarding the one loop continuum calculation. We would also like to 
thank Georg von Hippel and Peter Lepage for many helpful
discussions during the course of this project and Chris Bouchard for reading 
an early version of the manuscript.
 This work was supported in part by the
U.S. DOE, Grants No.~DE-FG02-04ER41302 and No.~DE-SC0011726. E.G.~is supported
in part by MINECO (Spain) under Grants No.~FPA2010-16696 and No.~FPA2006-05294; 
by Junta 
de Andaluc\'{\i}a
(Spain) under Grants No.~FQM-101 and No.~FQM-6552; and by the European 
Commission under
Grant No.~PCIG10-GA-2011-303781. Some of the
computing was undertaken on the Darwin supercomputer at the HPCS, University of
Cambridge, as part of the DiRAC facility jointly funded by the STFC, and on the
sporades cluster at the College of William and Mary.
\end{acknowledgments}

\appendix

\section{\label{app:oneloop}COMMENTS ON THE CONTINUUM ONE LOOP CALCULATION}

In this Appendix we give some details of the continuum one loop calculations 
entering the matching procedure. We focus mainly on scheme dependence, 
particularly in the ``SLL sector'', the sector that covers operators $Q2$ and 
$Q3$.  The continuum results given in Sec.~\ref{sec:contqcdres} appeared in
\cite{Gamiz:2008sk} and expressions for $c_{11}$, $c_{12}$, $c_{22}$ and 
$c_{21}$ were first published in \cite{Hashimoto:2000eh}. For those 
calculations the BBGLN scheme \cite{Beneke:1998sy} was used in the SLL sector.  
In an Appendix of \cite{Gamiz:2008sk} results were also presented in the SLL 
sector in the BMU scheme \cite{Buras:2000if}, another popular scheme, denoted 
$\tilde{c}_{22}$, $\tilde{c}_{21}$, $\tilde{c}_{33}$ and $\tilde{c}_{31}$. We 
have since discovered errors in results for $\tilde{c}_{33}$ and 
$\tilde{c}_{31}$ and correct them here.

We use the NDR-$\msb$ scheme to regularize ultraviolet divergences. We employ a 
gluon mass, $\lambda$, to handle infrared divergences, as in the lattice 
calculations. To fix a renormalization scheme completely within dimensional 
regularization of four-fermion operators, one must also  specify one's choices 
of evanescent operators, which enter the calculations as counterterms. 
Hence one starts from a specific basis of physical operators and then lists the 
evanescent operators that arise when one tries to project complicated Dirac 
structures in loop diagrams back onto the physical basis.  Most calculations in 
the literature follow the renormalization procedures with evanescent operators 
of Buras and Weisz \cite{Buras:1989xd}. For one loop calculations it 
is more convenient to list projections onto the physical basis for the various 
Dirac structures encountered. Then the evanescent operators are defined as 
the difference between left-hand and right-hand sides of these projection 
relations. The evanescent operators vanish in $d=4$ dimensions by construction, 
and for $d\neq 4$ dimensions they are understood to be subtracted away 
through the renormalization process. In the Buras and 
Weisz renormalization 
scheme \cite{Buras:1989xd}, equations explicitly involving evanescent operators 
will become relevant only at two loops. Even at one loop, however, and staying 
within the framework of the Buras and Weisz renormalization scheme, the set of 
evanescent operators is not unique. Different  projections correspond to 
different evanescent operators being subtracted by the renormalization 
procedure. Different  projections also lead to different finite contributions 
to the matching coefficients (the $c_{ij}$'s), although the one loop anomalous 
dimensions remain the same.

\subsection{Examples from the VLL sector}

Essentially all continuum calculations used in phenomenology are in agreement 
on the choices for evanescent operators relevant for $Q1$, $Q4$ and $Q5$. A 
well-known projection relation, for instance, in the $Q1$ sector (also called 
the ``VLL sector''), is given by
\begin{equation}
\label{projvll2}
[\gamma_\mu \gamma_\nu \gamma_\rho P_L \otimes 
\gamma^\mu \gamma^\nu \gamma^\rho P_L]   =  (16 - 2 \epsilon) \; [\gamma_\rho 
P_L \otimes \gamma^\rho P_L],
\end{equation}
where we use $d = 4 - \epsilon$ and $P_L \equiv 1 - \gamma_5$. Eq.~
\eqref{projvll2} is equivalent to defining and writing down the evanescent 
operator,
\begin{align}
\label{eopvll2}
E^{\mathrm{VLL}}_2 = {} & \left ( \overline{\Psi}_b^i \gamma_\mu \gamma_\nu 
\gamma_\rho 
P_L \Psi_q^i \right ) \, 
 \left ( \overline{\Psi}_b^j \gamma^\mu \gamma^\nu \gamma^\rho 
P_L \Psi_q^j \right ) \nonumber \\
{} & \qquad - (16 - 2 \epsilon) \; Q1 .
\end{align}
Another evanescent operator in the VLL sector is
\begin{equation}
\label{eopvll1}
E^{\mathrm{VLL}}_1 = \left ( \overline{\Psi}_b^i  \gamma_\rho 
P_L \Psi_q^j \right ) \, 
 \left ( \overline{\Psi}_b^j  \gamma^\rho 
P_L \Psi_q^i \right ) -  Q1 .
\end{equation}
In order to write the ``projection'' version of this definition we work with 
Dirac structures $[\Gamma_a \otimes \Gamma_b]$ sandwiched between external 
spinors. This allows us to take the different color contractions (e.g. ``iijj'' 
or ``ijji'') into account. In other words if,
\begin{equation}
\left\langle \left ( \overline{\Psi}_1^i  \Gamma_a 
 \Psi_2^i \right ) \, 
 \left ( \overline{\Psi}_3^j  \Gamma_b
 \Psi_4^j \right ) \right\rangle_{\mathrm{tree}} = [\overline{u}_1 \Gamma_a 
u_2] 
\, [\overline{u}_3 \Gamma_b u_4],
\end{equation}
then
\begin{equation}
\left\langle \left ( \overline{\Psi}_1^i  \Gamma_a 
 \Psi_2^j \right ) \, 
 \left ( \overline{\Psi}_3^j  \Gamma_b
 \Psi_4^i \right ) \right\rangle_{\mathrm{tree}} = - [\overline{u}_1 \Gamma_a 
u_4] 
\, [\overline{u}_3 \Gamma_b u_2].
\end{equation}
This step appears between Eqs.~\eqref{eq:q124tree} and \eqref{eq:q357tree} 
in the main 
text.
In this notation the projection 
version of Eq.~\eqref{eopvll1} becomes
\begin{align}
\label{projvll1}
[\overline{u}_1 \gamma_\rho P_L u_4] \, [\overline{u}_3 \gamma^\rho P_L 
u_2] = {} &  
- [\overline{u}_1 \gamma_\rho P_L u_2] \, [\overline{u}_3 \gamma^\rho P_L 
u_4] \nonumber \\
{} & \qquad  - \left\langle E_1^{\mathrm{VLL}} \right\rangle.
\end{align}
Again the 
operator $\langle E_1^{\mathrm{VLL}} \rangle$ is subtracted away in most
renormalization schemes and does not contribute in Eq.~\eqref{projvll1} 
(see 
Appendices 
A and B of Ref.~\cite{Buras:2012fs} that discuss this point). We have 
used projections such as (\ref{projvll2}) and (\ref{projvll1}) in deriving 
$c_{11}$ and $c_{12}$ of Sec.~\ref{sec:contqcdres}.

\subsection{The SLL sector in the BBGLN scheme}

We turn next to the SLL sector, which includes operators such as $Q2$ 
and $Q3$ and also, in some schemes, the tensor operator 
\begin{equation}
QT \equiv 
\left ( \overline{\Psi}_b^i  \sigma_{\mu \nu}  
P_L \Psi_q^i \right ) \, 
 \left ( \overline{\Psi}_b^j  \sigma^{\mu \nu}
P_L \Psi_q^j \right ),
\end{equation}
where $\sigma_{\mu \nu} = \frac{1}{2} [\gamma_\mu , \gamma_\nu]$. As mentioned 
earlier, our continuum results for $c_{22}$, $c_{21}$, $c_{33}$ and $c_{31}$ in 
Sec.~\ref{sec:contqcdres} are given in the ``BBGLN'' scheme, introduced in 
\cite{Beneke:1998sy}. This scheme uses $Q2$ and $Q3$ as the physical operator 
basis. Eq.~(15) of \cite{Beneke:1998sy} defines their evanescent operators
 in the SLL sector through the following projection:
\begin{align}
\label{projsll}
 [\overline{u}_1 \gamma_\mu \gamma_\nu P_L u_2]  {} &\,
[\overline{u}_3 \gamma^\mu \gamma^\nu P_L u_4] = \nonumber\\
{} & \qquad 2 (4 - \epsilon) 
\, 
 [\overline{u}_1  P_L u_2] \, [\overline{u}_3  P_L u_4]  \nonumber \\
{} & \qquad - 4(2 - \epsilon) \, 
 [\overline{u}_1  P_L u_4] \, [\overline{u}_3  P_L u_2] .
\end{align}
Equivalently one can list the evanescent operators 
\begin{align}
E^{\mathrm{SLL}}_1 = {}&  \left ( \overline{\Psi}_b^i  \gamma_\mu \gamma_\nu 
P_L \Psi_q^i \right ) \, 
 \left ( \overline{\Psi}_b^j  \gamma^\mu \gamma^\nu 
P_L \Psi_q^j \right ) \nonumber \\
 & \qquad   - 2 (4 - \epsilon) Q2 \; - \; 4 (2 - \epsilon)  Q3,
\end{align}
and
\begin{align}
E^{\mathrm{SLL}}_2 = {} &  \left ( \overline{\Psi}_b^i  \gamma_\mu \gamma_\nu 
P_L \Psi_q^j \right ) \, 
 \left ( \overline{\Psi}_b^j  \gamma^\mu \gamma^\nu 
P_L \Psi_q^i \right )   \nonumber \\
 & \qquad  -2 (4 - \epsilon) Q3 \;  - \; 4 (2 - \epsilon)  Q2 .
\end{align}
Using projections such as \eqref{projsll}, we first calculate the 
one loop corrections to $Q2$ and $Q3$, including the mixing between these two 
operators. This gives
\begin{equation}
\label{bbgln}
\left( \begin{array}{c}
   \langle Q2 \rangle \\
   \langle Q3 \rangle 
                   \end{array}  \right)_{\msb}
 = \left [ I + \alpha_s  \widehat{M} \right ]
\left( \begin{array}{c}
   \langle Q2 \rangle \\
   \langle Q3 \rangle 
                   \end{array}  \right)_{\mathrm{tree}}
\end{equation}
with
\begin{equation}
\widehat{M} = \left( \begin{array}{cc}
                 c^\prime_{22} \; \; c^\prime_{23} \\
                 c^\prime_{32} \; \; c^\prime_{33}
                              \end{array}  \right).
\end{equation}
We note that these are the full continuum QCD results, with external momenta 
$p_q = 0$ 
for the light quarks and $p_Q = (\pm M, \vec{0})$ for the heavy (anti)quarks.
The on-shell spinors obeying
 $\overline{u}_Q p^\mu \gamma_\mu = M \, \overline{u}_Q$ and   
 $\overline{v}_Q p^\mu \gamma_\mu = -  M \, \overline{v}_Q$ then  also 
obey 
 $\overline{u}_Q \gamma_0 =  \overline{u}_Q$ and   
 $\overline{v}_Q \gamma_0 =  - \overline{v}_Q$.
This allows us to use the large $M$ relation,
\begin{equation}
\label{r0}
\langle Q2 \rangle_{\mathrm{tree}} + \langle Q3 \rangle_{\mathrm{tree}} + 
\frac{1}{2} \langle Q1 \rangle_{\mathrm{tree}} = 0.
\end{equation}
So, the $c_{ij}$ in Sec.~\ref{sec:contqcdres} for the VLL+SLL sector become
\begin{align}
c_{22} = {} &  c^\prime_{22} - c^\prime_{23},  \qquad 
c_{21} =  - \frac{1}{2} c^\prime_{23},\\
c_{33} = {} &  c^\prime_{33} - c^\prime_{32},  \qquad
c_{31} =  - \frac{1}{2} c^\prime_{32}.
\end{align}

\subsection{The SLL sector in the BMU scheme}

The ``BMU'' scheme picks $Q2$ and $QT$ for the physical basis in the SLL sector.
Ref.~\cite{Buras:2012fs} presents a very convenient set 
of projections for this scheme in their Appendix B, which covers the full 
basis, 
$Q1$, $Q2$, $QT$, $Q4$ and $Q6$.  Here we reproduce just those for 
the SLL sector:
\begin{align}
[\gamma_\mu \gamma_\nu P_L \otimes {} & \gamma^\mu \gamma^ \nu P_L ]  = 
(4 - \epsilon) \, [P_L \otimes P_L] \nonumber \\
{}  & \qquad + [\sigma_{\mu \nu} P_L \otimes \sigma^{\mu 
\nu} P_L] \\
 [\gamma_\mu \gamma_\nu P_L \otimes {} &\gamma^\nu \gamma^\mu P_L 
 ]  =  (4 - \epsilon) \, [P_L \otimes P_L] \nonumber \\
{} & \qquad  - [\sigma_{\mu \nu} P_L \otimes \sigma^{\mu 
\nu} P_L] \\
[\sigma_{\mu \nu} \gamma_\alpha \gamma_\beta P_L \otimes {} &
\sigma^{\mu \nu} \gamma^\alpha \gamma^\beta P_L ]  = 
(48 - 40 \epsilon) \, [P_L \otimes P_L] \nonumber \\
 {}& \qquad + (12 - 3 \epsilon)[\sigma_{\mu \nu} P_L \otimes \sigma^{\mu \nu} 
P_L] \\
[\sigma_{\mu \nu} \gamma_\alpha \gamma_\beta P_L \otimes {} &
\gamma^\beta \gamma^\alpha \sigma^{\mu \nu} P_L ]  = 
- (48 - 40 \epsilon) \, [P_L \otimes P_L] \nonumber \\
{} & \qquad + (12 - 7 \epsilon) [\sigma_{\mu \nu} P_L
 \otimes \sigma^{\mu \nu} P_L] 
\end{align}
Note that since all five operators in this basis have the same color 
structure, we do not need to include external spinors in the projection 
relations. Instead of Eq.~\eqref{bbgln}, we now have 
\begin{equation}
\label{bmu}
\left( \begin{array}{c}
   \langle Q2 \rangle \\
   \langle QT \rangle 
                   \end{array}  \right)_{\msb}
 = \left [ I + \alpha_s \widehat{M}_{2T} \right ]
\left( \begin{array}{c}
   \langle Q2 \rangle \\
   \langle QT \rangle 
                   \end{array}  \right)_{\mathrm{tree}},
\end{equation}
with
\begin{equation}
\widehat{M}_{2T} = \left( \begin{array}{cc}
                 \tilde{c}^\prime_{22} \; \; \tilde{c}^\prime_{2T} \\
                 \tilde{c}^\prime_{T2} \; \; \tilde{c}^\prime_{TT}
                       \end{array}    \right).
\end{equation}
One can now rotate to the $Q2$, $Q3$ basis so that         
\begin{equation}
 \left( \begin{array}{cc}
   \langle Q2 \rangle \\
   \langle Q3 \rangle 
                   \end{array}  \right)_{\mathrm{tree}}
 = \widehat{R} 
 \left( \begin{array}{cc}
   \langle Q2 \rangle \\
   \langle QT \rangle 
                   \end{array}  \right)_{\mathrm{tree}}, \quad
\widehat{M}_{Q23} = \widehat{R} \widehat{M}_{2T} \widehat{R}^{-1},
\end{equation}
where
\begin{equation}
\widehat{R} = 
             \left( \begin{array}{cc}
                \; 1 \; \; 0 \\
                 - \frac{1}{2} \; \; \frac{1}{8}
                              \end{array}  \right).
\end{equation}
Finally we use Eq.~\eqref{r0} once again to obtain 
$\tilde{c}_{22}$,
$\tilde{c}_{21}$,
$\tilde{c}_{33}$,
$\tilde{c}_{31}$ in the BMU scheme. 
The updated and corrected results are
\begin{align}
\tilde{c}_{22}& = & \frac{1}{4 \pi} \left \{6 + \frac{16}{3} {\rm log}
\frac{\mu^2}{M^2} - \frac{4}{3} {\rm log}\frac{\lambda^2}{M^2} \right \} \\
\tilde{c}_{21}& = & \frac{1}{4 \pi} \left \{\frac{4}{3} + \frac{1}{3} {\rm log}
\frac{\mu^2}{M^2} + \frac{2}{3} {\rm log}\frac{\lambda^2}{M^2} \right \} \\
\tilde{c}_{33}& = & \frac{1}{4 \pi} \left \{- \frac{2}{3} - \frac{8}{3} {\rm 
log}
\frac{\mu^2}{M^2} - \frac{4}{3} {\rm log}\frac{\lambda^2}{M^2} \right \} \\
\tilde{c}_{31}& = & \frac{1}{4 \pi} \left \{\frac{17}{6} + \frac{4}{3} {\rm log}
\frac{\mu^2}{M^2} + \frac{2}{3} {\rm log}\frac{\lambda^2}{M^2} \right \} .
\end{align}
As expected, the anomalous dimension terms and infrared logarithms are 
the same as in the BBGLN scheme.

\bibliography{/home/cjm97/bibliography/bibliography.bib}

% End document
\end{document}